\documentclass[pre,twocolumn,preprintnumbers,showpacs,floatfix]{revtex4}
\usepackage{epsf}
\usepackage{graphicx}
\usepackage{psfig}
\usepackage{dcolumn}
\usepackage{bm}
\usepackage{subfigure}
\usepackage{color}
\usepackage{amsthm}
\usepackage{amssymb}
\usepackage{amsmath}
\usepackage{pb-diagram}
\usepackage{algorithm}
\usepackage{algorithmic}

\newcommand{\fn}{{\mathsf{FN}}}
\newcommand{\cn}{{\mathsf{CN}}}
\newcommand{\pd}{{\mathsf{PD}}}

\newcommand{\setof}[1]{\left\{ {#1}\right\}}

\newcommand{\R}{{\mathbb{R}}}

\def\setof#1{\left\{{#1}\right\}}
\def\ang#1{\langle {#1} \rangle}

\definecolor{gray85}{gray}{0.85} 
\definecolor{gray8}{gray}{0.8} 
\definecolor{gray7}{gray}{0.7} 
\definecolor{gray6}{gray}{0.6} 
\definecolor{gray5}{gray}{0.5} 
\definecolor{gray4}{gray}{0.4} 
\definecolor{gray35}{gray}{0.35} 

\begin{document}

\title{
Evolution of Force Networks in Dense Particulate Media
}

\author{ Miroslav Kram\' ar}
\email{miroslav@math.rutgers.edu}
\author{Arnaud Goullet}
\email{arnaud.goullet@gmail.com}
\affiliation{Department of Mathematics,
Hill Center-Busch Campus,
Rutgers University,
110 Frelinghusen Rd,
Piscataway, NJ  08854-8019, USA}
\author{Lou Kondic}
\email{kondic@njit.edu}
\affiliation{Department of Mathematical Sciences,
New Jersey Institute of Technology,
University Heights,
Newark, NJ 07102}
\author{Konstantin Mischaikow}
\email{mischaik@math.rutgers.edu}
\affiliation{Department of Mathematics and BioMaPS Institute,
Hill Center-Busch Campus,
Rutgers University,
110 Frelinghusen Rd,
Piscataway, NJ  08854-8019, USA}

\begin{abstract}
We introduce novel sets of measures with the goal of describing  dynamical properties of force networks in dense 
particulate systems.  The presented approach is based on persistent homology and allows for extracting
precise, quantitative measures that describe  the evolution of geometric features of the interparticle forces, without necessarily considering the details related to individual contacts between particles.   
The networks considered emerge from 
discrete element  simulations of two dimensional particulate systems consisting of compressible frictional circular disks. 
We quantify the evolution of the networks for slowly compressed systems undergoing jamming transition.   The main 
findings include uncovering significant but localized changes of force networks for
unjammed systems, 
global (system-wide) changes as the systems evolve through jamming, to 
be followed by significantly less dramatic evolution for the jammed states.   We consider
both connected components, related in loose sense to force chains, and loops, and find that both measures provide a significant insight into the evolution
of force networks.  In addition to normal, we consider also tangential forces between the 
particles and find that they evolve in the consistent manner.  Consideration
of both frictional and frictionless systems leads us to the conclusion that 
friction plays a significant role in determining the dynamical properties of the considered networks.
We find that the proposed approach describes the considered networks in a precise yet tractable manner, allowing to 
identify novel features which could be difficult or impossible to describe using other approaches.  
\end{abstract}


\maketitle

\section{Introduction}

Particulate systems have been extensively studied through centuries due to their importance to our 
everyday life.   These systems appear everywhere, from nano to cosmic scales, and may evolve
either hard particles (such as sand) or soft ones (emulsions, foams), see, e.g.~\cite{liu95,brujic03,cates98,majmudar05a}.
As these system are exposed to some 
external (e.g. compression) or internal (electric, magnetic, gravitational...) influence they may 
compress, reaching a stage that the particles are in more-or-less permanent contacts.    As 
the systems evolve in time (for whatever reason), the contacts between them in general evolve as well.

Formation of contact networks between the particles, and their properties, have been extensively 
studied in many different contexts, and using a number of different  tools including 
percolation and network type of approaches; see~\cite{albert_barabasi_02, alexander_physrep05} 
for overviews.    In addition to contact networks, however, there is 
an additional network of interactions, often called force network in the literature related
to granular materials.   These force networks are not simply slaved to contact
networks, due to indeterminacy of the interaction between particles.   A simple example of this 
is the fact that  multiple force networks may be consistent with the condition of force and 
torque balance in a system.   These force networks include complete information about a
system and therefore is of significant interest to describe and eventually understand their properties, 
in particular since it is well known that the interparticle forces play a key role in determining the mechanical 
properties of static and dynamic systems; see e.g.~\cite{alexander_physrep05} for an extensive review in the
context of amorphous solids.

Physical systems of relevance typically consist of large number of particles, and therefore the 
force networks may become extremely complex.   Due to this complexity, it is necessary
to develop techniques that lead to an understanding of the important properties of these networks,
without necessarily considering all the details, since this would lead to an intractable study.
One obvious idea is to consider statistical properties of these networks, and to ask,
e.g., what is the probability of having a force between two particles of a given magnitude. Such
studies have been carried out for  granular and other systems  (see, e.g.,~\cite{radjai98b,majmudar05a,silbert_pre06})
and have led to a significant new insight.  Even on the statistical level, however, there are still open
questions - one example is a recent discussion of the probability of presence of large forces in 
granular systems~\cite{majmudar05a,corwin05}.  

Going beyond purely statistical description requires 
analyzing in more detail the local properties of the force networks.  Such studies have been extensively
utilized only recently, but have already indicated the complexity of the problem.   Examples include
detailed discussion of the forces between particles, see~\cite{peters05,tordesillas_pre10,tordesillas_bob_pre12} and
the references therein, where these local properties were connected to global response of the systems
considered.     Going beyond statistical level, and in the same time attempting to keep the focus on the global
structure of the interaction networks requires implementation of new techniques.   Considered approaches include
network-type of analysis~\cite{daniels_pre12, herrera_pre11,walker_pre12}.
These works provide a significant new insight and confirm that the properties of force networks are relevant in the 
context of propagation of acoustic signals~\cite{daniels_pre12}, fracture~\cite{herrera_pre11}, and compression and 
shear~\cite{walker_pre12}.  Topology based approach has been also considered, with focus on the contact network topology in
isotropically compressed~\cite{arevalo_pre10}  and tapped granular media~\cite{arevalo_pre13}. 

We have recently started employing 
algebraic topological techniques for the purpose of quantifying forces in a manner which is global in character,
but at the same time includes detailed information about the geometric structures of the forces.
In~\cite{epl12} we analyzed the number of components
and loops, measured by Betti numbers (to be described in more detail below), as a function of 
compression (packing fraction) and the force level. That work considered the force networks
on particle level: essentially, a total force on each particle was computed, and then the features of this 
force field were considered by analyzing the number of clusters (components), related in a broad 
sense to so-called force chains, as well as the number of loops, related to force chains' connectivity.
Realizing that a more complete description of a system can be reached by 
considering particle-particle interactions explicitly, we turned our attention to force
networks whose basic building blocks are the interaction forces at particle contacts.   
Governed still by the idea of considering global properties, we have used persistent homology \cite{edelsbrunner:harer}, which is 
best thought of a continuous map that provides for a substantial reduction of data while preserving geometric structures.  
The output of persistent homology consists of persistence diagrams, $\pd$'s, that provide a consistent identification of the force thresholds at 
which geometric features made up of sets of interacting particles appear and disappear.
Persistence analysis has shown explicitly and on a global level that (i) the geometry of the forces between particles has 
different properties from the geometry of the contact networks and (ii) the properties of these forces may depend 
strongly on the material properties,  such as friction or polydispersity, for the specific case of granular particles~\cite{pre13}.

Due to the complexities involved in the studies of force networks based on any of the approaches
discussed so far, most of the existing results have concentrated
on analysis of these networks for static systems, and there are very few attempts to analyze
dynamical aspects of the force networks, or even of time-dependent properties of the 
forces experienced by particles, see e.g.,~\cite{corwin_pre08} and the 
references therein.    Clearly,  dynamical aspects are of significant importance, since many  of the systems of interest 
are time dependent, and one would like to understand how the networks evolve
in such a setting; some recent examples in the literature where the evolution of 
interaction networks is clearly of importance include wave propagation through
particulate systems~\cite{daniels_pre12} and impact~\cite{clark_prl12}.

There are many questions related to temporal evolution of force networks that one could ask, 
such as: What are the generic features of the temporal evolution of force networks?  Are these
features different for unjammed, but dense systems compared to the jammed ones?   What
happens as a system goes through jamming transition?    What is the influence of friction on 
the evolution of force networks?  We will address some of these questions in the present work. 
The tools that we will use  involve the concept of `distance' between $\pd$'s, 
measuring (in a manner that will be made precise) the amount of change in force networks from one state to the next.   
To be specific, we will consider a particular system of inelastic frictional particles, but the 
technique that will be described is independent of the model describing particle-particle interaction,
and could be equally well applied to the systems of particles of different shapes interacting by any other means.   

The remainder of this paper is organized as follows. In Sec.~\ref{sec:system} we discuss 
the system to be studied.  A brief introduction to  persistence is given in Sec.~\ref{sec:methods}; more in-depth 
analysis is given in~\cite{physicaD13}.  In this section we also introduce the concept of distance between 
the persistence diagrams.    In Sec.~\ref{sec:results} we present the results of persistence 
analysis.  Section~\ref{sec:conclusions} summarizes the results and discusses possible future directions.

\section{The systems to be studied}
\label{sec:system}

We focus on the system of soft inelastic possibly frictional disks in two spatial dimensions (2D), bound by inward moving rough
walls, similarly as in our earlier works~\cite{epl12,pre13}.     The simulations are of the type 
utilized in~\cite{epl12}; for completeness, a brief outline is given in Appendix~\ref{app:DES},
and a short description follows.  

The inward motion of wall particles is slow as described in the Appendix,
so that the energy provided by the walls' motion is dissipated quickly and therefore compression
waves or other type of large scale inhomogeneities are not present in the system: the packing fraction, $\rho$, is 
essentially uniform, except possibly for lower $\rho$'s, which are not the focus of this study.
The particles are characterized by the coefficient of Coulomb friction, $\mu$, and polydisperse, 
with a random distribution of diameters in the 
range $1\pm r_p/2$ scaled by the average particle diameter, $d_{ave}$; we use $r_p  = 0.4$.    
The initial size of the domain considered is  $L^2$, with $L = 50 d_{ave}$.  The initial configuration
is formed by placing particles on the rectangular lattice and then giving them initial random 
velocities.   After this initially supplied energy is dissipated (through
friction and inelasticity), the system is compressed leading to $\rho$'s in the range $[0.63:0.90]$. 
Gravitational effects are not considered.   

\begin{figure*}[t]
\begin{picture}(400,400)
\put(-60,230){\includegraphics[width=2.2in]{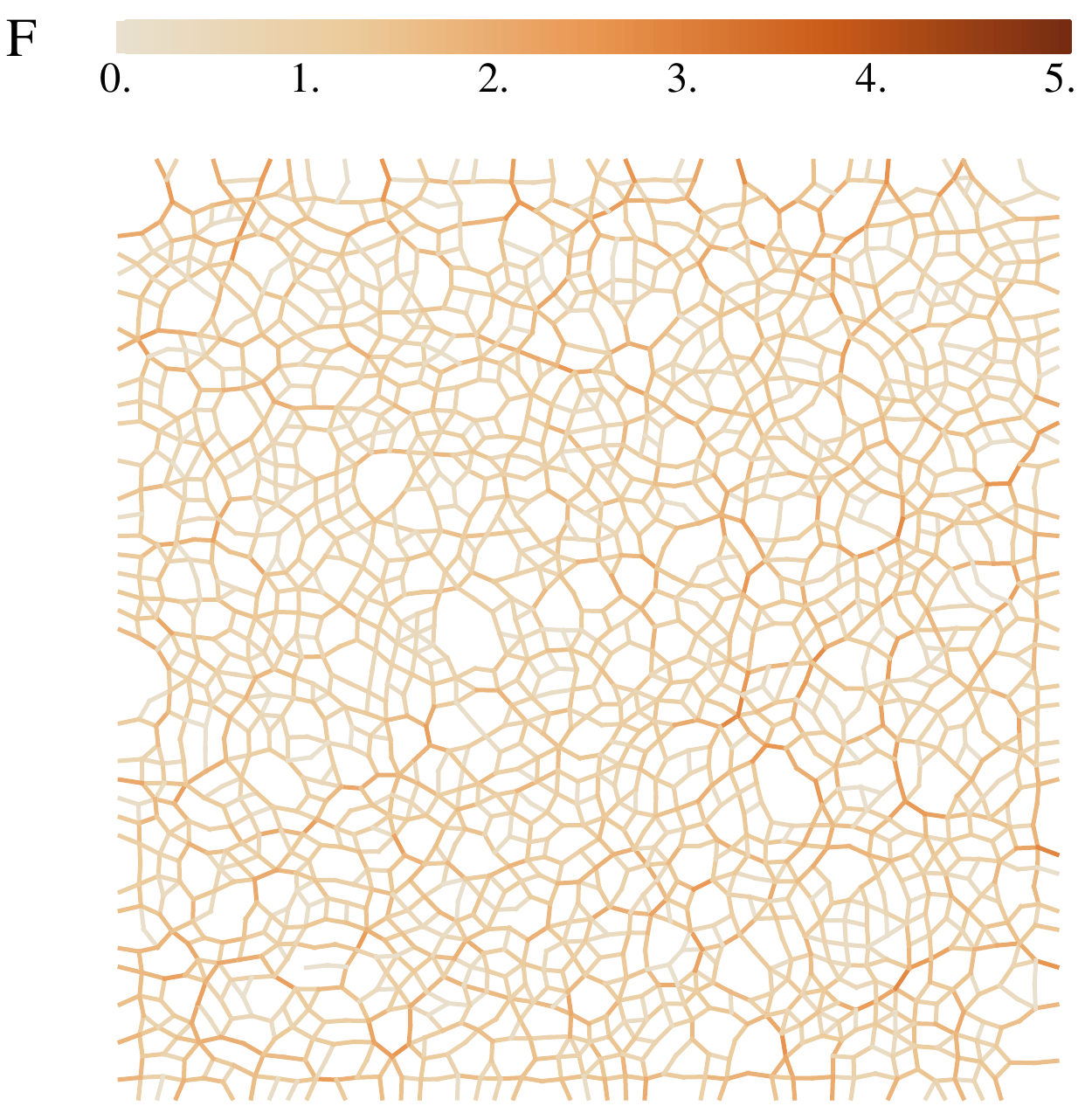}}
\put(105,230){\includegraphics[width=2.2in]{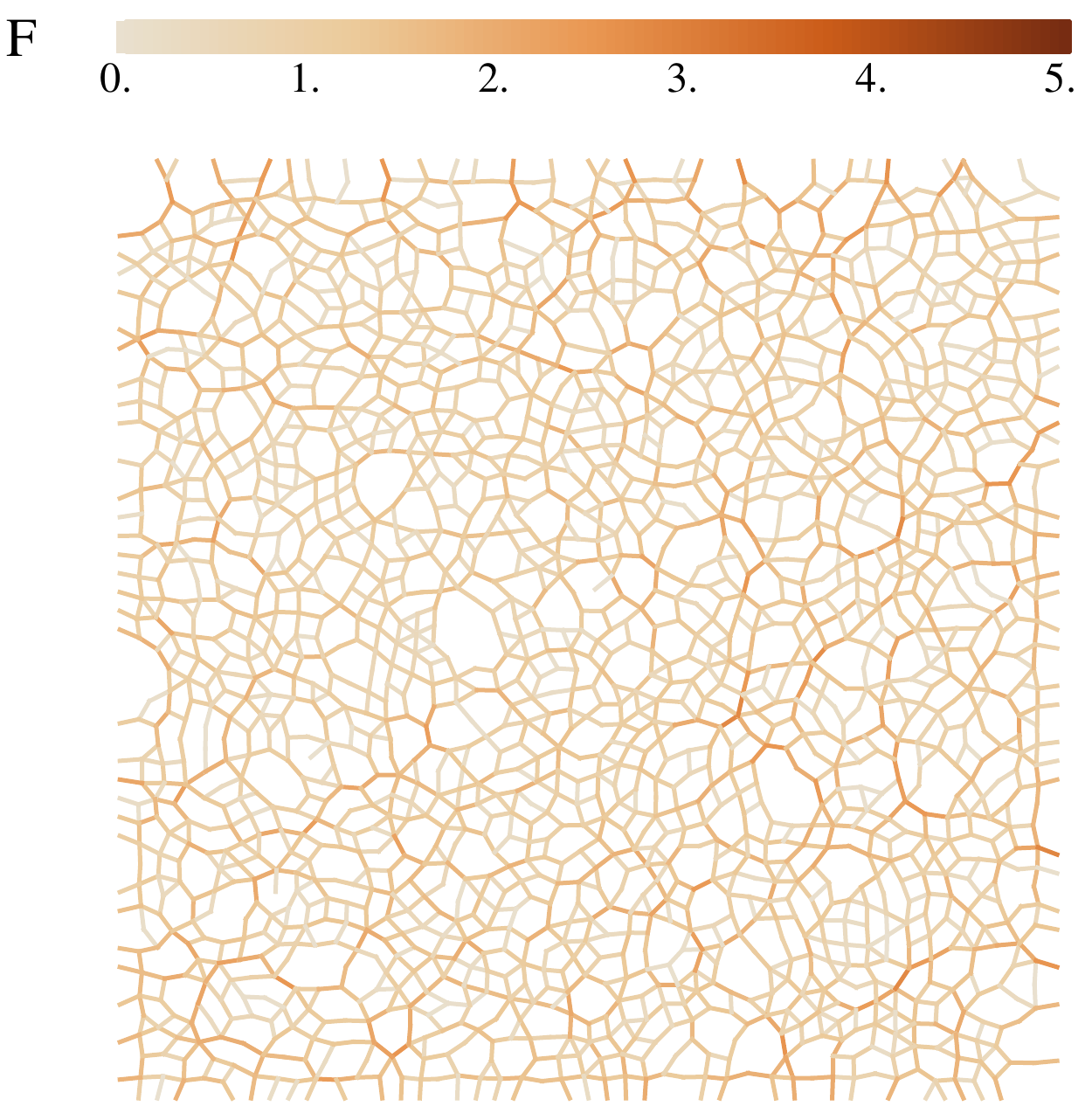}}
\put(270,230){\includegraphics[width=2.2in]{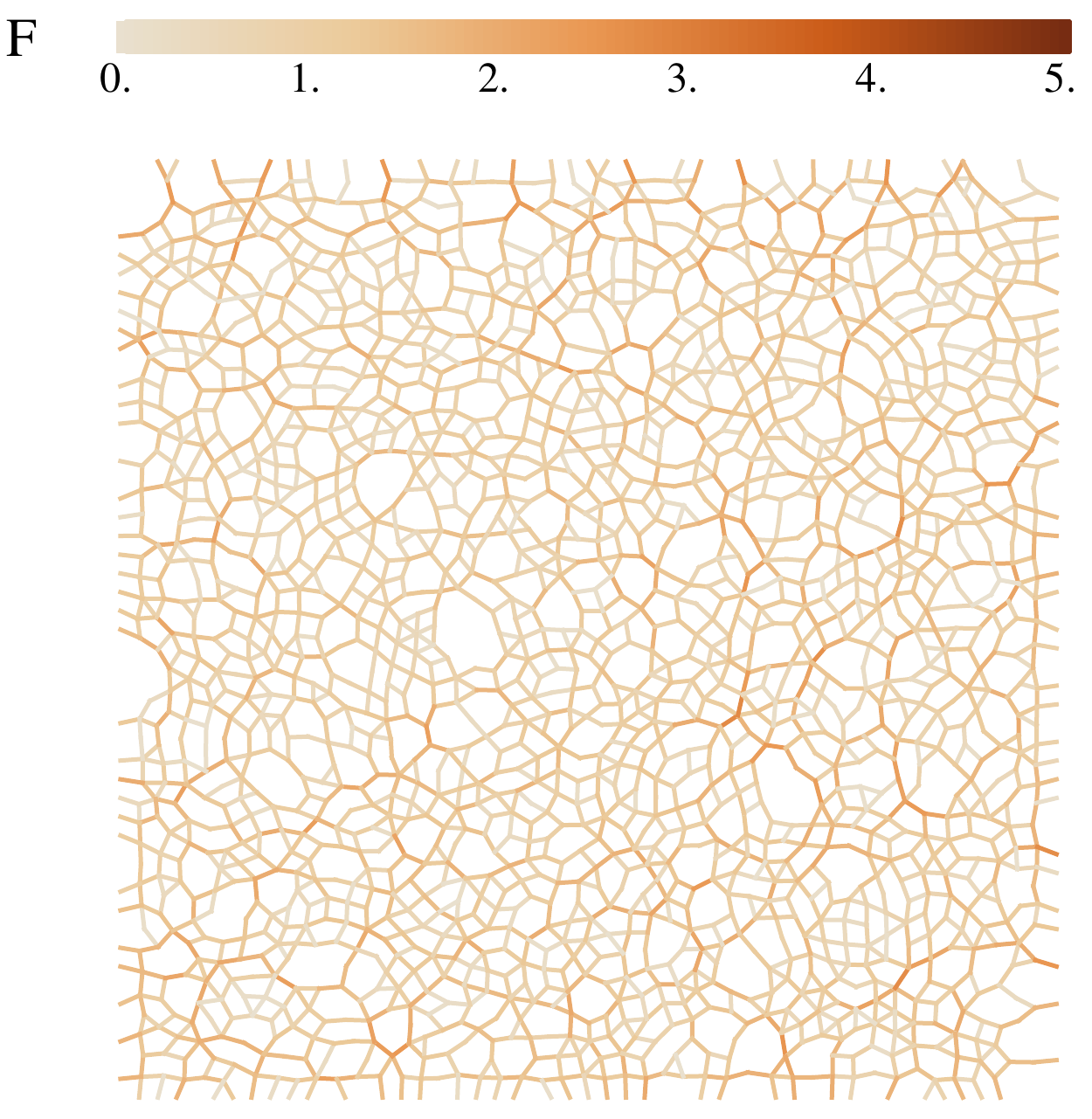}}
\put(-40,-5){\includegraphics[width=3in]{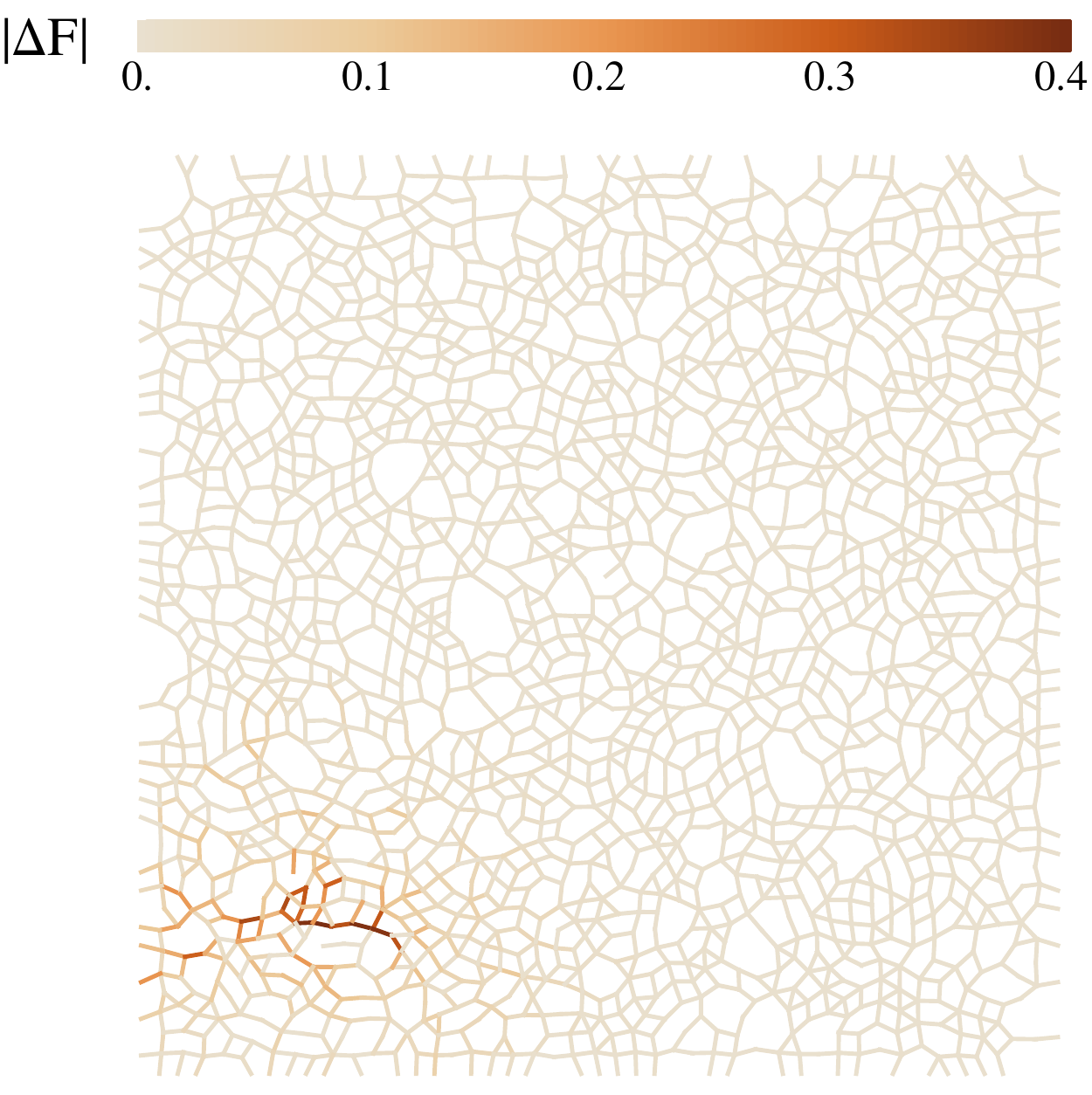}}
\put(210, -5){\includegraphics[width=3in]{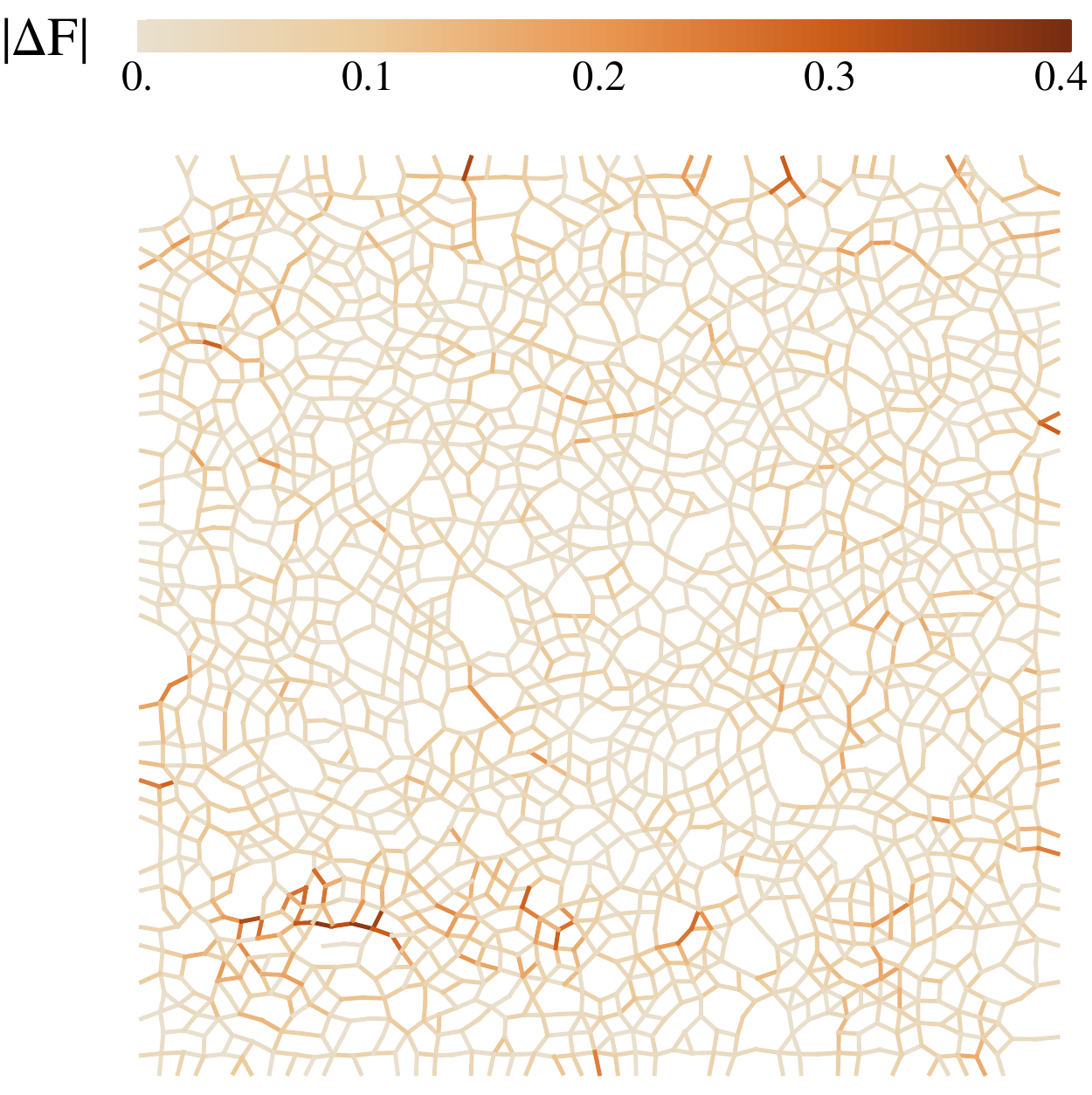}}

\put(20,220){(a)}
\put(180,220){(b)}
\put(350,220){(c)}
\put(65,-10){(d)}
\put(320,-10){(e)}
\end{picture}
\caption{(a-c) Snapshots of the simulated system ($r_p = 0.4~\mu=0.5$) at  three values of $\rho$ close to  $0.79$. Every edge in the network corresponds to 
the normal force acting between the particles and the color encodes the magnitude of the normal force normalized 
by the current value of the average normal force. 
(d-e)) The absolute values of the differences between the normal forces shown in (a-b) and (a-c), respectively (note different color scheme compared to (a-c).
}
\label{fig:ThreeNetworks}
\end{figure*}
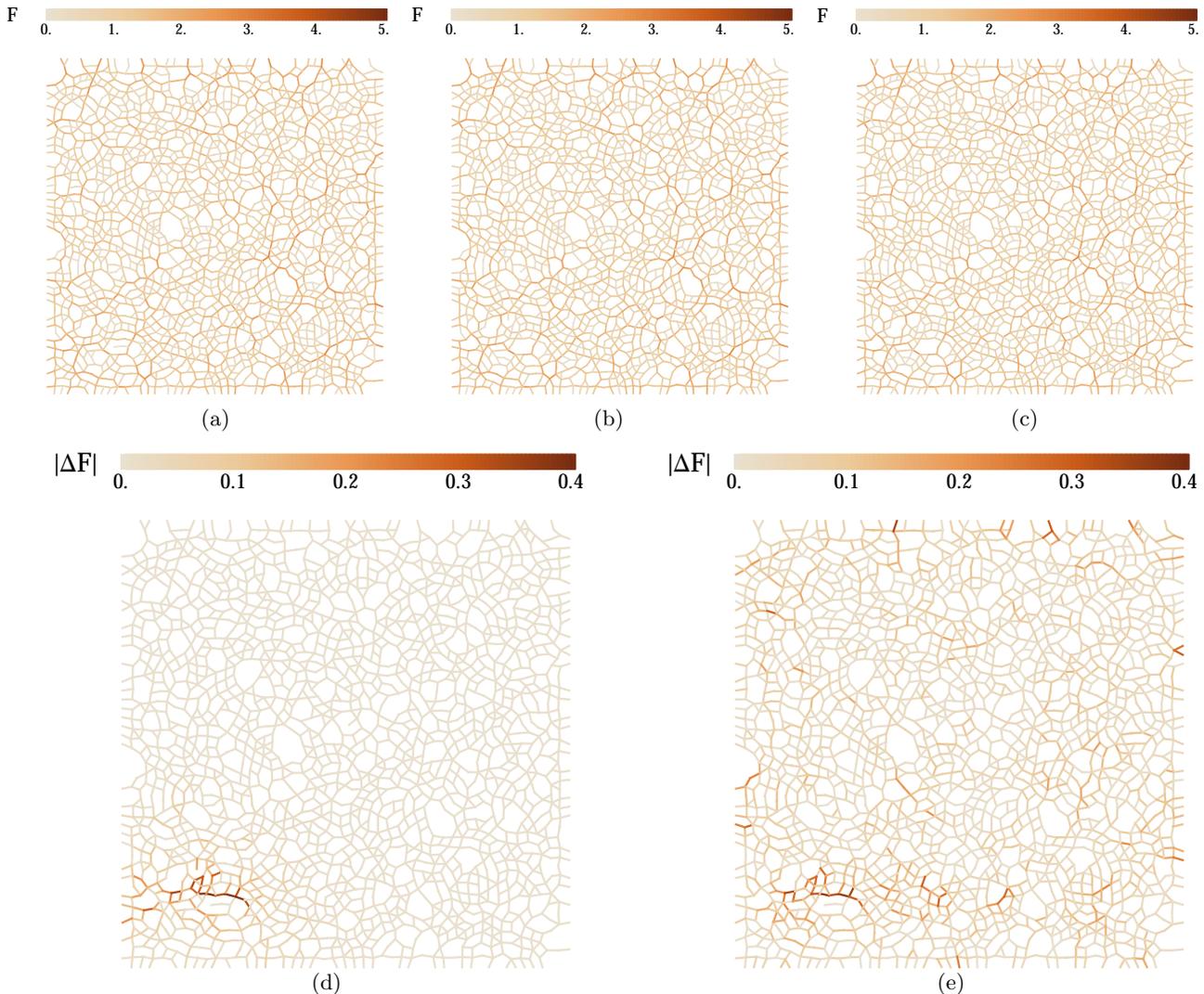

As the system is compressed, we extract the current values of the forces
between the particles at specified time intervals.   To provide visual picture of the system that is being analyzed,
Fig.~\ref{fig:ThreeNetworks}(a-c)  shows the magnitude of the normal forces acting between the particles for three nearby 
values of $\rho$ (animations of the evolution of the force network as  the system is compressed are available as 
Supplementary Materials of~\cite{pre13}).
The forces are normalized by the average normal force at the given time. Note that the figures 
appear very similar, and additional inspection shows that both the average contact number, $Z$, and the average 
force are almost identical.   Differences between the networks  become  visible if we plot changes in  the 
magnitude of the normal force.  Figure~\ref{fig:ThreeNetworks}(d-e) shows these differences, and we can 
see that the difference between (a-b) is localized, while the (a-c) differences have more of a global character.   The main focus
of this paper is to introduce and discuss a set of measures that can be used to quantify these differences in a precise manner.

\section{Methods}
\label{sec:methods}

We start by describing force networks in the manner that will allow us to use topological techniques for 
their quantification.   As we will see,  important features of  the force network can be captured by 
using persistence diagrams,  $\pd$'s, that were already introduced in our previous work~\cite{pre13}, and discussed
in depth in~\cite{physicaD13}.    A brief overview of persistence is given in Sec.~\ref{sec:ForceNetwork}, 
followed by a couple of  examples in Sec.~\ref{sec:toy}, and discussion of $\pd$'s for the 
simulation data in Sec.~\ref{sec:diagrams}.  Then, in Sec.~\ref{sec:distance}
we discuss different types of metric on the space of $\pd$'s. The main feature of these metrics  is different 
sensitivity to various changes of the force network. This makes them useful for distinguishing local and global changes, as 
we demonstrate  by analyzing the networks shown in Fig.~\ref{fig:ThreeNetworks}.    The computations that
are outlined in this Section and used for the rest of the paper are based on publicly available open source 
software~\cite{mischaikow:nanda:11,perseus}.

\subsection{Force Networks}
\label{sec:ForceNetwork}

We encourage the reader to view the images shown in Fig.~\ref{fig:ThreeNetworks} as 2D landscapes with the 
altitude given by the normal force magnitude at a particular spatial point. The goal of this section is to briefly recall
the basic elements of persistent homology which we use to provide a concise characterization of the geometry of these landscapes.

We begin by making precise the process by which we obtain Fig.~\ref{fig:ThreeNetworks}(a - c).
Given a collection of particles $\setof{p_i\mid i=0,\ldots, N}$, we define a simplicial complex $\cn_I$
called the {\em interaction complex} consisting of vertices
$\setof{v_i\mid i=0,\ldots, N}$  where each vertex $v_i$ is identified with particle $p_i$ and
all possible edges $\ang{v_i,v_j}$. 
Now let $\psi_{i,j}\in\R$ denote the magnitude of the force experienced between particles $p_i$ and $p_j$, 
then the function  $f$ is defined on the edges by
\[
f(\ang{v_i, v_j}) := \psi_{i,j}.
\]
This function is extended to the vertices by 
\[
f(\ang{v_i}) = \max_{j=0,\ldots, N} \setof{f(\ang{v_i,v_j})}. 
\]
Figure~\ref{fig:ThreeNetworks}(a - c) indicates the value of $f$ on all edges $\ang{v_i,v_j}$ for which
$f(\ang{v_i,v_j})>0$.

Observe that in these figures there are triangles made up of three edges. These represent loops
in the contact  network  made up of three particles.  Since in a perfect densely packed crystalline 
structure  made up of disks of the same size all loops would be made up of exactly $3$ particles, we 
refer to a loop involving  four or more particles  as a {\em defect}.
In the analysis we perform in this paper we have 
chosen to focus on defects. To avoid counting the three particle loops we extend
$\cn_I$ to its flag complex $\cn^{\blacktriangle}_I$  by adding all triangles $\ang{v_i,v_j,v_k}$
and defining 
\[
f(\ang{v_i,v_j,v_k}) = \min \setof{f(\ang{v_i, v_j}),f(\ang{v_i, v_k}),f(\ang{v_j, v_k})}.
\]

Since our goal is to consider networks of contacts with forces larger than a given force, 
we are interested in the geometry of a part of the complex on which the forces exceed a specified level.
In particular, given a threshold $\theta >0$, define the associated {\em interaction force network}
\begin{equation}
\label{eq:superlevel}
 \fn_I(f,\theta) := \{ \sigma \in \cn^{\blacktriangle}_I \mid  f(\sigma) \geq \theta\},
\end{equation}
which corresponds to the part of the  contact  network experiencing force larger than $\theta$.  
Since we are working with finite sets of particles, the function $f$ can take on a finite set of values
$\Theta = \setof{\theta_n}$. The {\em interaction force network filtration} is the collection of 
interaction force networks
$\setof{\fn_I(f,\theta)\mid \theta \in\Theta}$.
For simplicity we drop the adjective interaction and write force network and force network filtration.

\subsection{Persistent Homology: an example}
\label{sec:toy}

\begin{figure*}
\centering
{\includegraphics[width=3.1in]{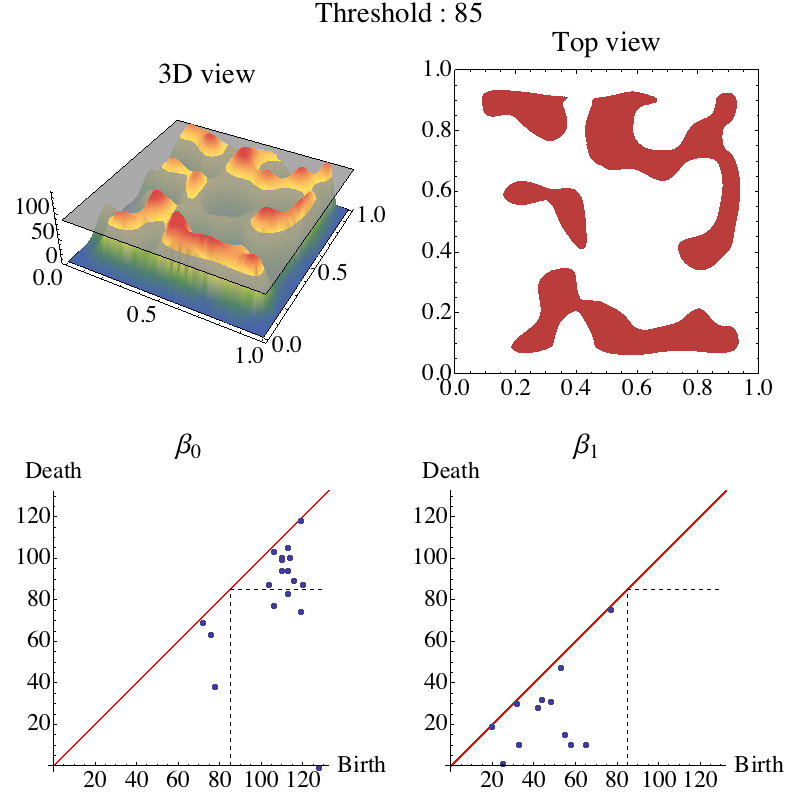}}  
{\includegraphics[width = 3.1in]{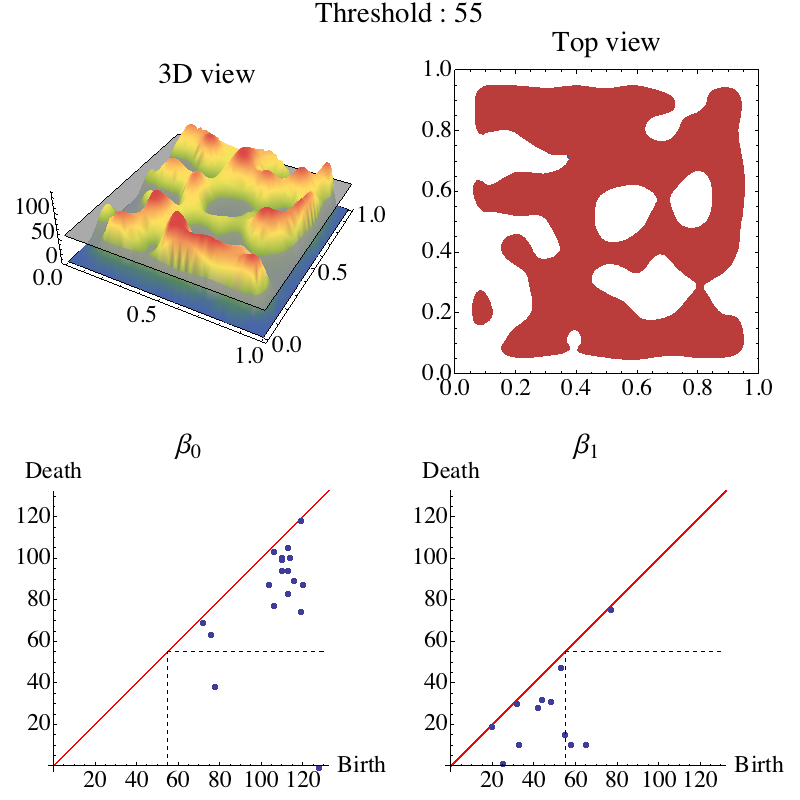}}
\caption{2D example illustrating $\pd$'s for a function of two variables.  For each threshold level we 
show the 3D view as well as the top view where only super levels are visible.   The  dashed lines in the $\pd$'s 
for $\beta_0$ (components) and $\beta_1$ (holes)  show the current threshold level.
Complete animation can be found in Supplementary Materials~\cite{sup_mat}. 
}
\label{fig:2D_toy}
\end{figure*}

In the context of the two dimensional simulations that are studied in this paper the simplest geometric structures 
that can be quantified by persistent homology are connected components and loops.
We use the simple two-dimensional example shown in Fig.~\ref{fig:2D_toy} to provide some intuition
to what these measurements represent.  The reader is referred to \cite{physicaD13} for a more complete
description, along with a simpler one-dimensional example.
Figure~\ref{fig:2D_toy} shows two threshold values, $85$ (a) and $55$ (b). 
The function (along with the associated threshold level) for which 
persistent homology is being computed is shown in the upper left hand corner of (a) and (b).  
The upper right hand corners show the associated the force networks, $\fn_I(f,85)$ and $\fn_I(f,55)$.
Each panel contains a $\beta_0$ $\pd$, denoted by $\pd_0$, which measures connected
components, and a $\beta_1$ $\pd$, denoted by $\pd_1$, which measures loops. Note that the $\pd$'s
in (a, b) are the same.

Since we are using super level sets to define the force networks, we
compute the $\pd$'s by descending through the threshold levels.
For the moment let us focus on Fig.~\ref{fig:2D_toy}(a) associated with the threshold $\theta = 85$.
The force network $\fn_I(f,85)$ contains no loops.  This can be seen in $\pd_1$ 
by noting that there are no points in the diagram to the lower right of $(85,85)$.  There
are, however, four components and in the $\beta_0$ diagram to the lower
right of $(85,85)$ there are four points: $(128,-1)$, $(119,74)$, $(113,83)$, and $(106,76)$.
We remark (this information is not contained in the $\pd_1$) that the point 
$(128,-1)$ corresponds to the lower component, 
$(119,74)$ to the upper right component, 
$(113,83)$ to the upper left component, and  
$(106,76)$ to the middle component.
These points contain important information: the first ({\em birth}) coordinate indicates the threshold at
which the component appears and second ({\em death}) coordinate indicates where the component merges
with another older component.  In particular, we can now conclude that the first point in the lower component
appeared at threshold level $128$, the first point in the upper right component at threshold level $119$,
etc.

To explain the comment about the "older component"  note that first component to 
appear appears at threshold level $128$.  Since this is the first component it is always the oldest
and thus never disappears. This is indicated by assigning a death coordinate of $-1$.
The death coordinate of the upper left component is $83$ which implies that at threshold $83$ it
merges with another component. The geometry suggests that it does not merge with the lower component
at that threshold. Thus, it can merge with the upper right component or the middle component.
However, the birth value of the middle component is $106$ which makes it younger than the 
upper left component. The upper right component has birth value $119$ which makes it older.
Thus when the upper right component merges with the upper left component at threshold $83$,
the upper left component dies.

Turning to Fig.~\ref{fig:2D_toy}(b) associated with threshold $\theta = 55$, we note that
to the lower right of $(55,55)$ there are two points in the $\pd_0$ 
and three points in the $\pd_1$, indicating that the associated force
network $\fn_I(f,55)$ has two components and three loops, respectively.
Observe that the lower loop has just formed and thus the associated birth coordinate is $55$.
The death value occurs at the threshold where a loop is filled in.

The convention that the younger feature dies has an important implication. 
Let $(\theta_b,\theta_d)$ be a point in a $\pd$. 
The number $\theta_b- \theta_d$ is called the lifespan of the geometric feature associated with  $(\theta_b,\theta_d)$. 
Observe that features with a longer lifespan persist over a longer range of values and hence are more robust. 
Conversely features with very short lifespans are often regarded as noise, since they
persist only over a small range of force values and thus can be introduced by small perturbations.

Viewing the complete animation of Fig.~\ref{fig:2D_toy} provided in the Supplementary Materials~\cite{sup_mat}
should convince the reader that persistent homology provides a concise encoding of the
dominant geometric features of a function.  The fact that it is concise implies that information
is lost and thus two distinct functions can have the same persistence diagram.  Of course,
if two functions have distinct persistence diagrams, then they must exhibit distinct geometric features.

\subsection{Persistence diagrams for simulation data}
\label{sec:diagrams}

\begin{figure}[thb]
\centering
\subfigure[Fig.~\ref{fig:ThreeNetworks}(a), $\beta_0$]{\includegraphics[width=1.6in]{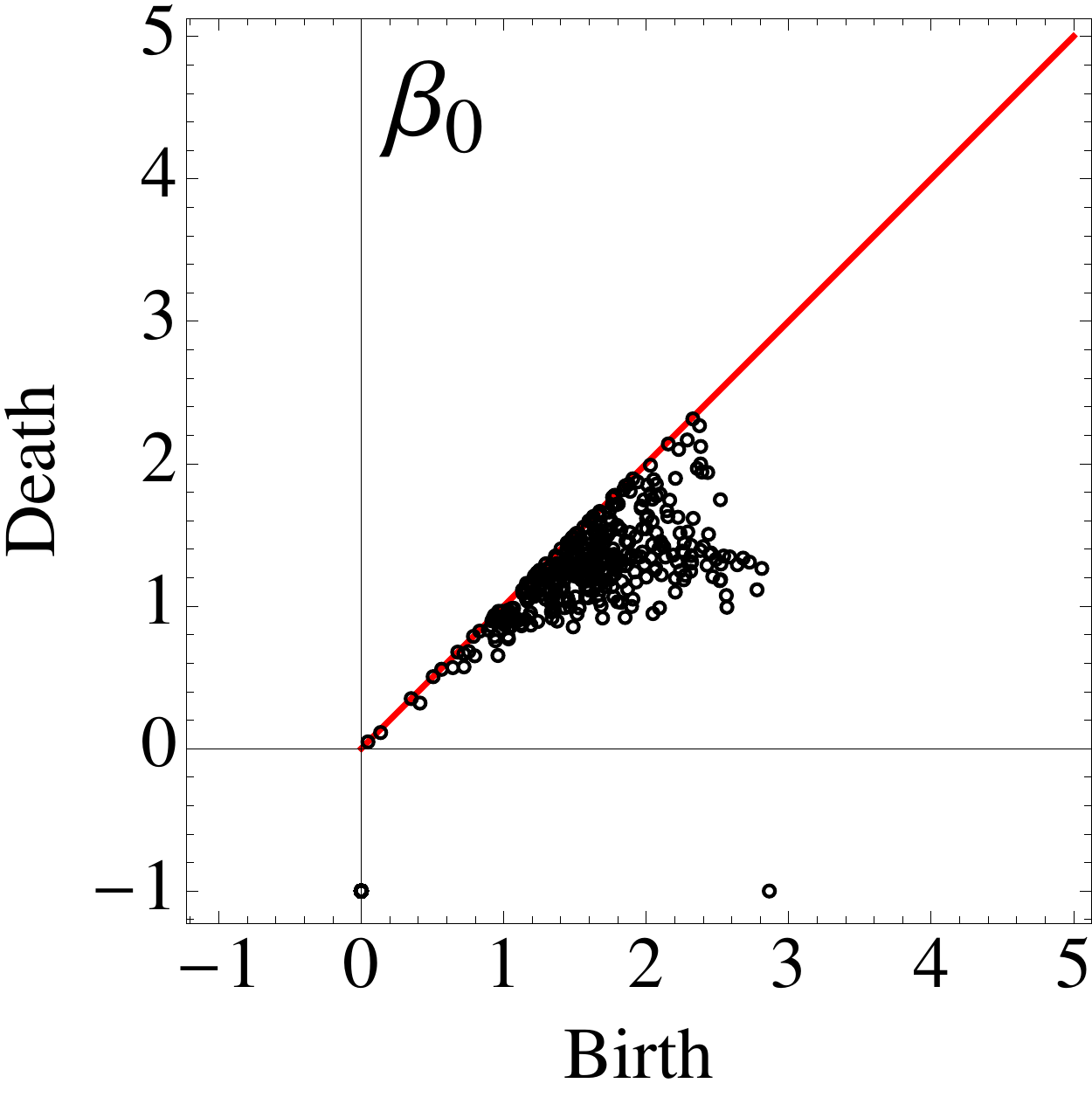}} 
\subfigure[Fig.~\ref{fig:ThreeNetworks}(a), $\beta_1$]{\includegraphics[width=1.6in]{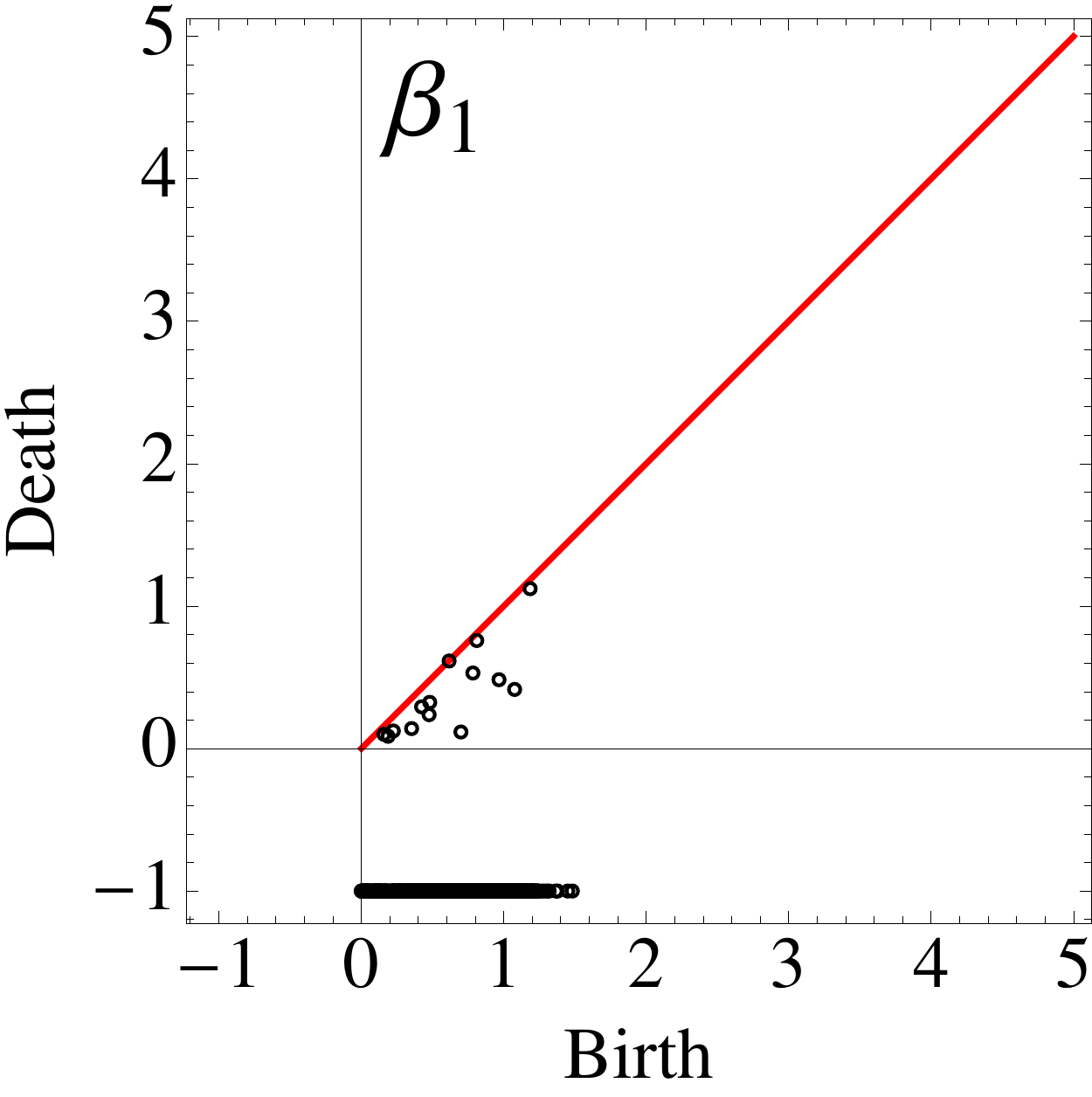}}\\
\caption{Persistence diagrams for the  interaction networks shown in Fig.~\ref{fig:ThreeNetworks}(a).  The diagrams 
corresponding to Fig.~\ref{fig:ThreeNetworks}(b - c) visually appear very similar (not shown).   
}
\label{fig:pers}
\end{figure}

Having discussed the simple example of the previous section, we are now equipped to consider $\pd$'s resulting from DES.
Figure~\ref{fig:pers} shows the $\pd_0$ and $\pd_1$ diagrams for the force network  defined on the flag complexes  corresponding to 
Figs.~\ref{fig:ThreeNetworks}(a) (the diagrams corresponding to Figs.~\ref{fig:ThreeNetworks}(b - c) visually appear very similar and are not
shown since they do not provide any  additional information).  Not surprisingly, these are more complex than those of Fig.~\ref{fig:2D_toy}.
We begin our analysis with some simple observations (see~\cite{pre13} and~\cite{physicaD13} for 
a more detailed discussion). 
\\
$\bullet$
The $\pd_0$ show a `cloud' of points in the $[0:3]$ birth range, meaning
that most of the features (force chains, loosely speaking) start appearing at the force level which is about $3$ times
the average force, and disappear by $[0.8:1]$, 
suggesting that at a force level slightly smaller than 
the average force, most features disappear (merge); 
\\
$\bullet$
Careful inspection of $\pd_0$ shows two points with a death value of $-1$ level; one born at high force threshold $\theta\approx 3$, and the other at the zeroth force level.   The interpretation of the former one follows from the convention discussed in the previous section that the component that is born first dies last; 
the latter is due to the presence of isolated particles (rattlers) 
or particles that experience only contacts with zero force.  Rattlers can be detected because only the higher dimensional  simplicies for which the function $f$ is positive are used in the persistent homology computations. Hence the rattlers create separate connected components at the zero force level.
\\
$\bullet$ 
Note that for $\pd_1$'s  a death level of $\theta_d = -1$
implies the existence of a defect.  If $\theta_d > 0$, then the hole is filled in, and 
$\theta_d$ indicates the weakest magnitude of interacting forces within the region enclosed by the loop.
The $\pd_1$ in  Fig.~\ref{fig:pers}(b) suggest that the birth value $\theta_b$ of most of the loops  are less 
than about $1.5$ the average force. 
That the birth values of the $\pd_1$ is lower than that of the $\pd_0$
is not surprising; the birth of a loop corresponds to the lowest magnitude of the normal forces
acting  along the contacts forming the loop. It is also worth noting that most of the loops are associated with defects.

While these observations about individual $\pd$'s are of interest, it is clear that a more systematic approach 
is needed to quantify and interpret the available information.    This is especially true in the context
of comparing time dependent systems, where each time step can produce a distinct $\pd$.  To make this comparison precise
requires the introduction of appropriate metrics to measure the difference between $\pd$'s.

\subsection{Distance between persistence diagrams}
\label{sec:distance}

\begin{figure}[t]
\centering
	\begin{picture}(400,115)
	\put(10,0){\includegraphics[width=2.0in]{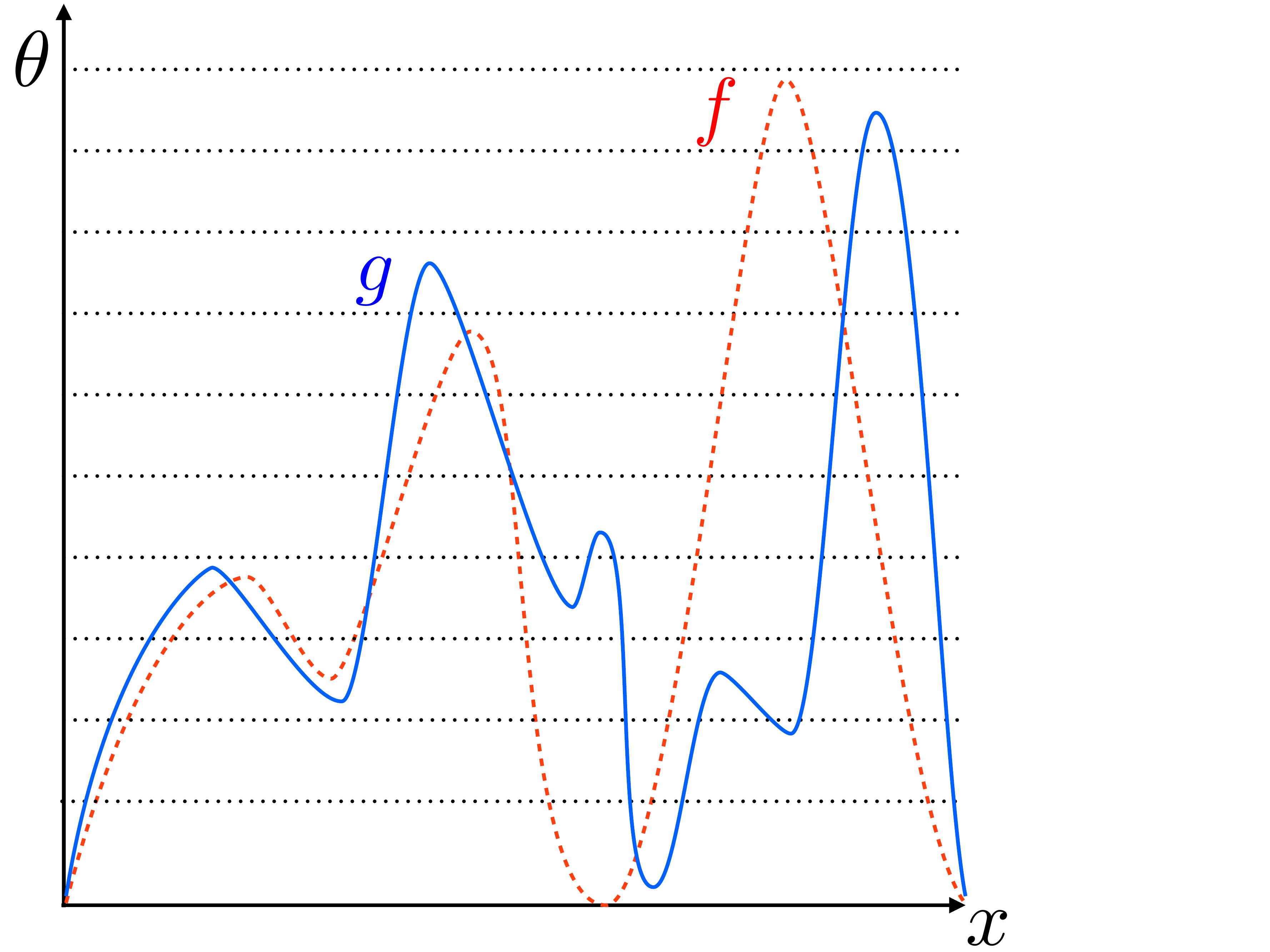}}
	\put(50,-5){(a)}
   	\put(135,-2){\includegraphics[width=2.0in]{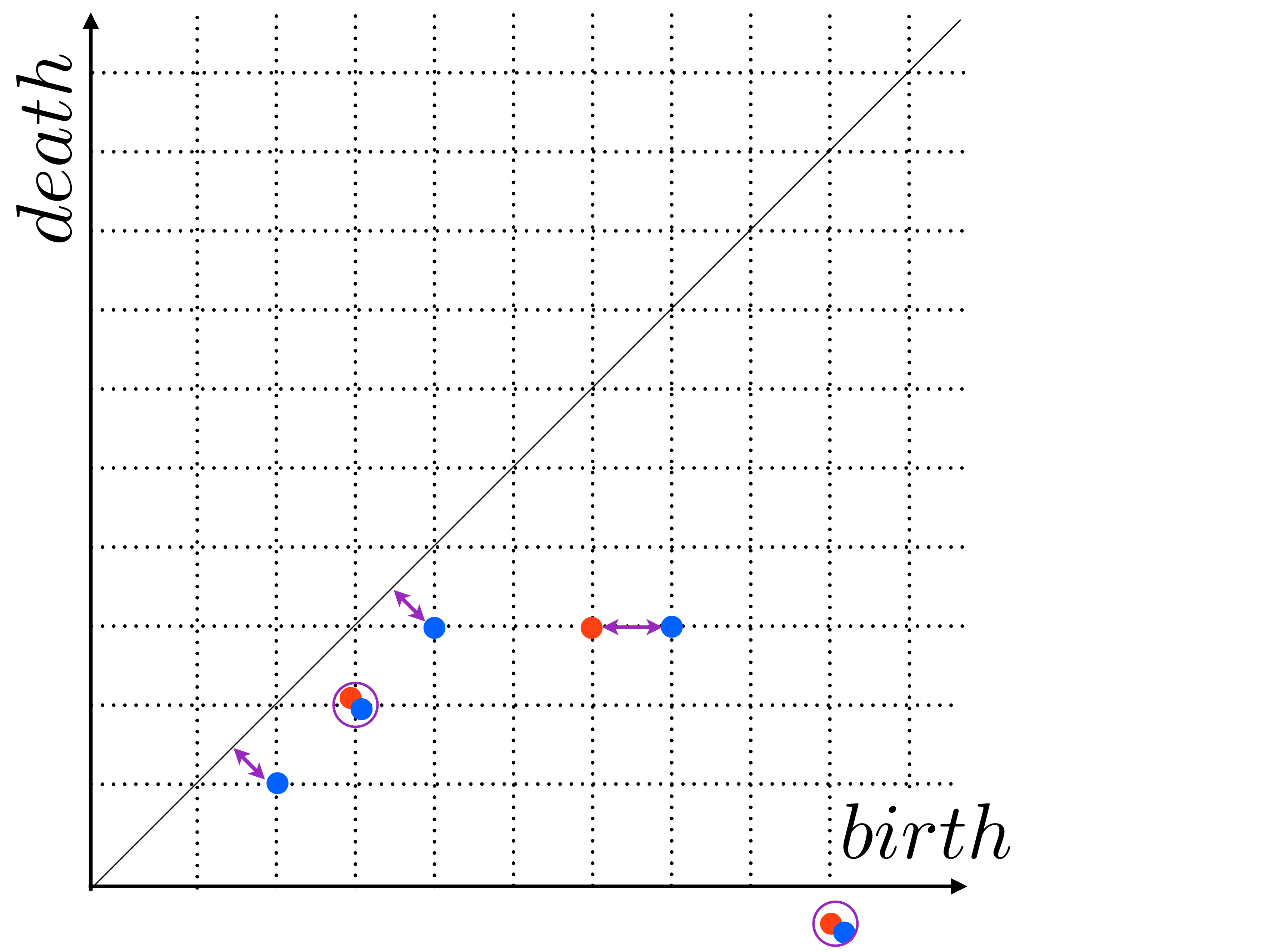}}
	\put(180,-5){(b)}
	\end{picture}
\caption{(a) Two functions. Blue represents a noisy perturbation of red. (b) Associated persistence diagrams along
with pairing of persistence points.  The Euclidean distances between these points are used to compute corresponding 
distances between the persistence diagrams as explained in the text.
}
\label{fig:bottleneck}
\end{figure}

At a minimum, an appropriate metric for $\pd$'s  must satisfy the property that if two functions are 
similar, then the distance between the associated $\pd$'s must be small.  It was shown in \cite{edelsbrunner:harer} that such metrics exist. Explicit formulas and a detailed discussion in the context of DGM is presented in 
\cite[Definition 7.1]{physicaD13}. However, since the precise definitions are somewhat technical, 
we limit ourselves to a heuristic presentation.  Consider Fig.~\ref{fig:bottleneck}(a) where one can 
consider the function $g$ as a noisy perturbation of $f$.  To understand the metrics we recall
two observations from  Sections~\ref{sec:toy} and \ref{sec:diagrams}: (1) if the 
points in two $\pd$'s lie in the similar regions then the $\pd$'s should be close,
and (2) short lifespans, i.e.\ persistence points near the diagonal, are related to small perturbations.
This suggests that given two $\pd$'s one attempts to match points from one diagram
with points in the other diagram  or points on the diagonal in such a way as to minimize  distances
between the  matched  points. Figure~\ref{fig:bottleneck}(a)
suggests such a  minimizing  matching. Let $\gamma\colon \pd_n(f) \to \pd_n(g)$, denote such a  matching
between two $\pd$'s. The {\em  bottleneck distance} is denoted by
$d_B(\pd_n(f),\pd_n(g))$ and is defined by  $\sup_{p\in \pd_n(f)} \| p -\gamma(p)\|_\infty$, while  the {\em degree-$q$ Wasserstein distance} is denoted by $d_{W^q}(\pd_n(f),\pd_n(g))$
and is defined by $\left(\sum_{p\in \pd_n(f)}\| p -\gamma(p)\|_\infty^q\right)^{1/q}$
where in both cases $\gamma$ is chosen to minimize these quantities. 

In this paper we restrict ourselves to the bottleneck distance $d_B$ and the  Wasserstein distances $d_{W^q}$,
$q=1,2$.
Observe that the bottleneck distance reports only the single largest difference between $\pd$'s
while $d_{W^q}$ includes all differences between the diagrams. Thus, it is always true that
\[
d_B \leq d_{W^q}.
\]
Sensitivity of the Wasserstein distances to small differences
(possibly due to noise) can be modulated by the choice of the value of $q$, i.e.\ $d_{W^2}$ is less sensitive
to small changes than $d_{W^1}$.

{Example shown in  Fig.~\ref{fig:BVsW}(a) demonstrates 
the differences between $d_B$ and $d_{W1}$.  Let $\pd_0(f)$ and $\pd_0(g)$ be the $\pd_0$'s
corresponding to the functions $f$ and $g$, respectively.
Fig.~\ref{fig:BVsW}(b) shows that  the  $\pd_0(f)$ consist of a single point $(5,-1)$. 
The $\pd_0(g)$ is more complicated. There is one copy 
of $(3,-1)$ corresponding  to the dominant feature of $g$ and  seven copies of $(1,0)$ representing  seven smaller features.  Since the
bottleneck distance considers only the largest feature, $d_B(\pd_0(f),\pd_0(g)) = 2$.   On the other hand, 
$d_{W^1}(\pd_0(f),\pd_0(g)) = 5.5$ since the contribution of every small feature to the $d_{W^1}$ distance is $0.5$ and $d_{W^2}(\pd_0(f),\pd_0(g)) \approx 2.6$. Thus the different metrics 
distances provide complementary information about the differences between two functions (or two landscapes).
In particular, the fact that the  $d_{W^2}$  distance is closer to  $d_B$  than $d_{W^1}$ suggests that
the geometric difference between $f$ and $g$ lies in the noise as opposed to the dominant features.

\begin{figure}
\centering
\begin{picture}(400,115)
	\put(-10,5){\includegraphics[width=2.2in]{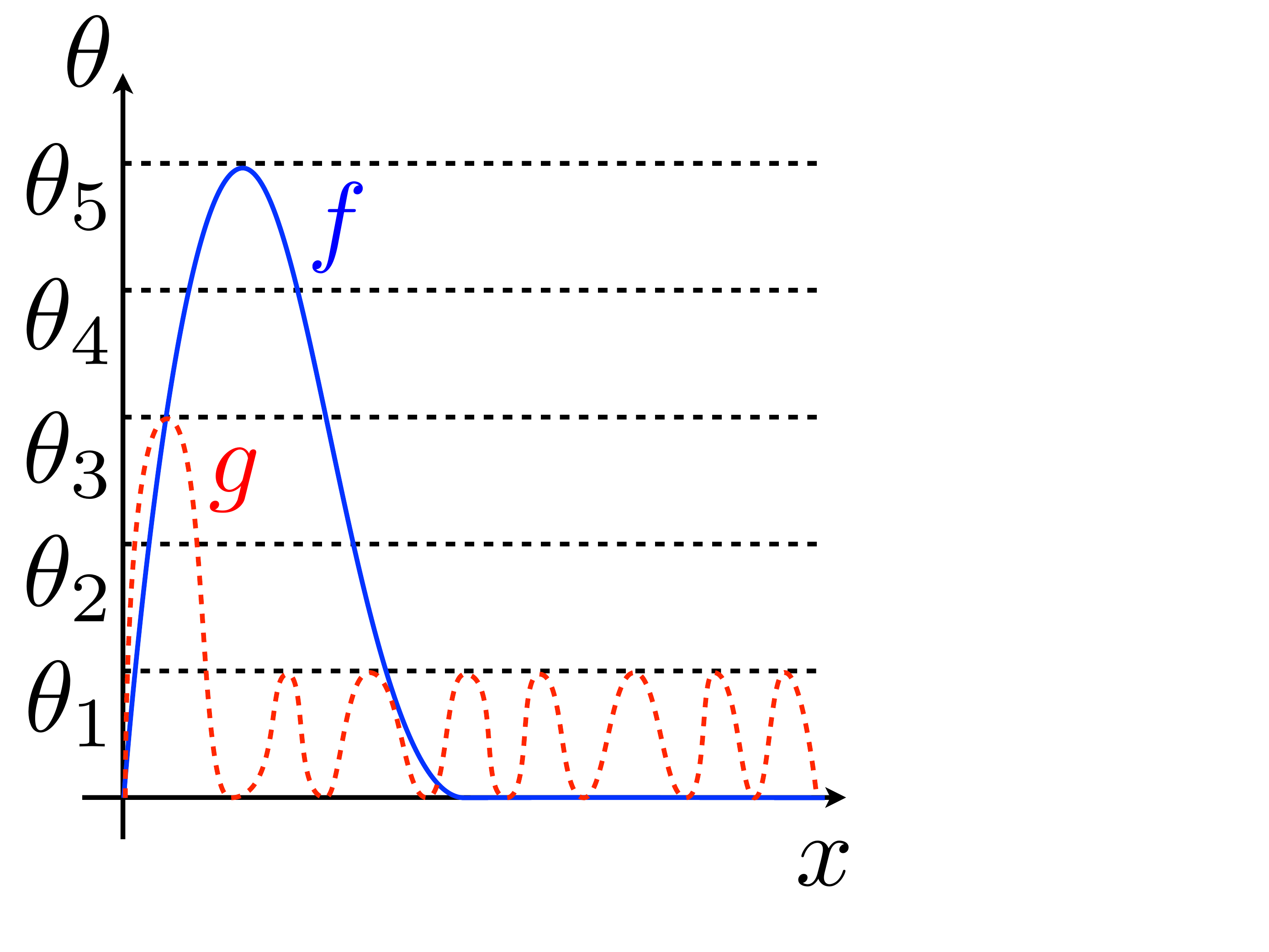}}
	\put(50,-5){(a)}
   	\put(105,5){\includegraphics[width=2.2in]{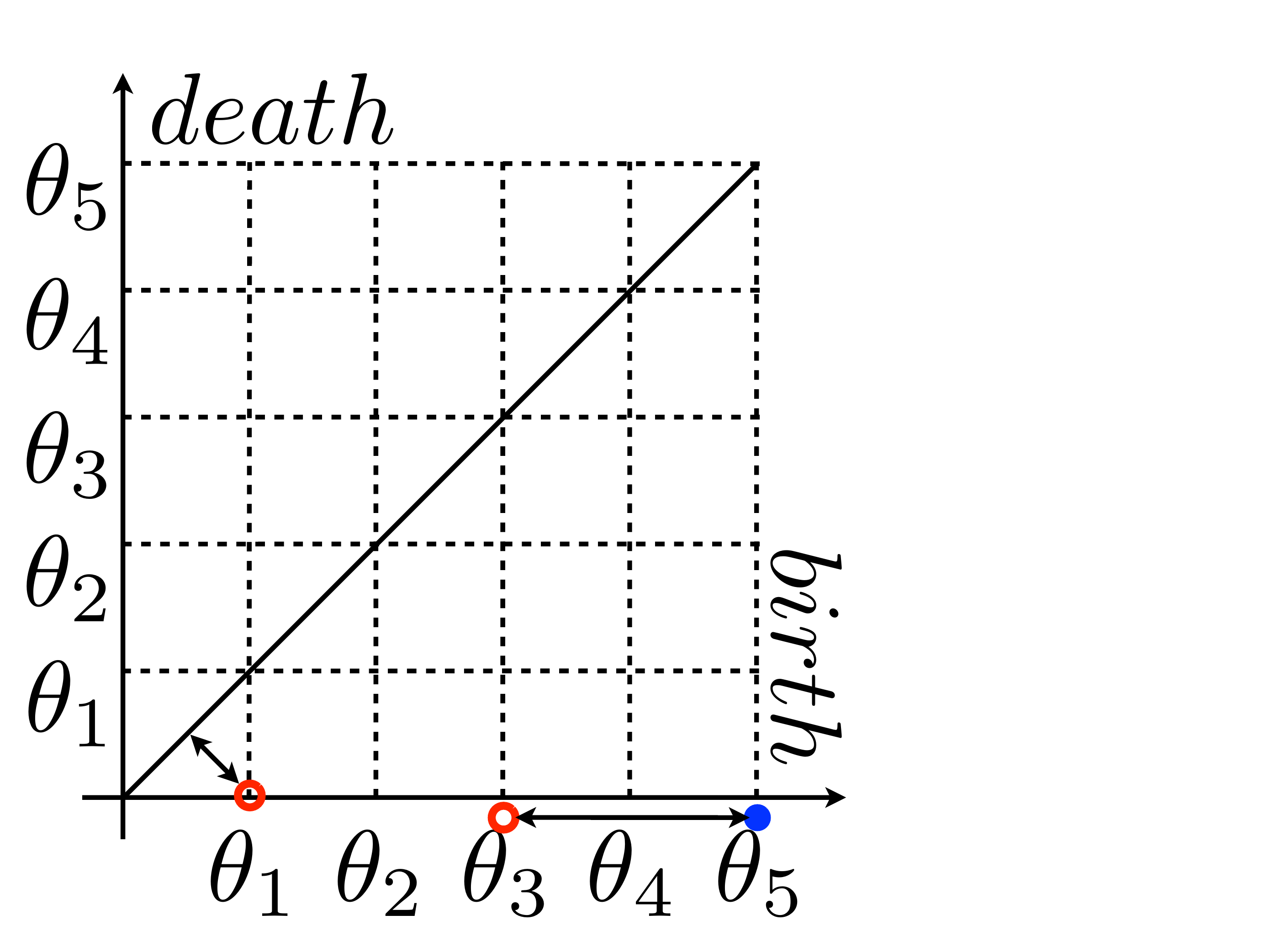}}
	\put(180,-5){(b)}
\end{picture}
\caption{(a) Two scalar fields $f$ (solid blue) and $g$ (dashed red). (b) Persistence diagrams for both scalar fields. The red hollow dot 
at $(\theta_1,\theta_0)$ has multiplicity seven because there are seven geometric features corresponding to the oscillations of $g$. 
The double arrows show the pairing corresponding to the bijection $\gamma$ for which the bottleneck  and 
 Wasserstein distance $d_{W1}$ are obtained.
}
\label{fig:BVsW}
\end{figure}

We now apply these different distances to the analysis of the forces shown in Fig.~\ref{fig:ThreeNetworks}(a-c).  
Let $\pd^a, \pd^b$ and $\pd^c$ denote the corresponding $\pd$'s (Fig.~\ref{fig:pers} shows the ones corresponding 
to the  part (a)).
Table~\ref{table:ThreeNetworks} gives the numerical values for the considered distances.
Observe that $d_{B}(\pd_i^a,\pd_i^b)$ and  $d_{B}(\pd_i^a,\pd_i^c)$, $i=0,1$, are small (less than 
10\% and 15\% of the average force, respectively).  Thus, there is no single dramatic change in the geometries
 between these landscapes.
 Note that this does not imply that there are no significant point wise differences
 between the force landscapes. In fact the range of Fig.~\ref{fig:ThreeNetworks}(d-e), that measures the point wise
 difference,  extends to $0.4$ ($40\%$ of the average force). Combining these two observations suggests that the point wise 
 locations of the strong and weak particle interactions have shifted, but their relative geometries have remained similar.
 
As is to be expected the values of the Wasserstein distances are larger than those of the bottleneck.
What is worth noting is that the relative difference between $d_{W^1}(\pd_i^a,\pd_i^c)$ 
and  $d_{W^1}(\pd_i^a,\pd_i^b)$ is significantly larger than the relative difference between 
 $d_{W^2}(\pd_i^a,\pd_i^c)$ and  $d_{W^2}(\pd_i^a,\pd_i^b)$.
This suggests that the force landscapes of Fig.~\ref{fig:ThreeNetworks}(c) differs from the one
of Fig.~\ref{fig:ThreeNetworks}(a) via many more 
small geometric changes  than the force landscape of
Fig.~\ref{fig:ThreeNetworks}(b) differs from that of Fig.~\ref{fig:ThreeNetworks}(a).
This is consistent with the difference plots of Fig.~\ref{fig:ThreeNetworks}(d-e) which show that in 
general the values of Fig.~\ref{fig:ThreeNetworks}(e) are higher than those of Fig.~\ref{fig:ThreeNetworks}(d)
and  spread over a broader range of the domain.

\begin{table}
\centering 
\begin{tabular}{c |c c c | c c c }
\multicolumn{1}{c}{} & \multicolumn{3}{c}{$\beta_0$} &   \multicolumn{3}{c} {$\beta_1$}\\
\hline & $d_B$  & $d_{W^1}$ & $d_{W^2}$  & $d_B$    & $d_{W^1}$ & $d_{W^2}$  \\
\hline $(\pd^a,~\pd^b)$  & $0.089$  & $4.18$  & $0.50$ & $0.081$   & $3.8$ & $0.57$  \\
\hline $(\pd^a,~\pd^c)$  & $0.12$  & $14.0$ & $0.89$  & $0.15$    & $19.2$ & $0.99$  \\
\end{tabular}
\caption{Distances between the persistence diagrams for the force networks 
corresponding to Fig.~\ref{fig:ThreeNetworks}(a-c). 
}
\label{table:ThreeNetworks}
\end{table}

So far, we have shown that the various distances can be used not only to quantify 
the differences between force networks, but also to distinguish between local and global differences;
we will use these findings to analyze the results of DES in the rest of this paper.  

Before concluding 
this section, we illustrate that the distances can be also used to isolate a part of the domain where  dominant
differences are present, although we will not use this approach in the rest of the paper.

\begin{table}
\centering 
\begin{tabular}{c | c c c | c c c }
\multicolumn{1}{c}{} & \multicolumn{3}{c}{ $\beta_0$} &   \multicolumn{3}{c} {$\beta_1$}\\
\hline$(\pd^a,~\pd^b)$ & $d_B$  & $d_{W1}$ & $d_{W2}$  & $d_B$    & $d_{W1}$ & $d_{W2}$  \\
\hline bottom left   & $0.12$  & $3.6$  & $0.50$ & $0.16$   & $3.66$ & $0.95$  \\
\hline bottom right & $0.014$  & $0.40$  & $0.070$ & $0.014$   & $0.43$ & $0.037$  \\
\hline top right      & $0.0046$  & $0.58$  & $0.35$ & $0.0055$   & $0.29$ & $0.024$  \\
\hline top left        & $0.0076$  & $0.56$ & $0.31$  & $0.0045$    & $0.23 $ & $0.019$  \\
\end{tabular}
\caption{Distances between the persistence diagrams for the different parts of the force networks corresponding to 
Fig.~\ref{fig:ThreeNetworks}(a,b).}
\label{table:AB}
\end{table}

\begin{table}
\centering 
\begin{tabular}{c|c c c | c c c }
\multicolumn{1}{c}{} & \multicolumn{3}{c}{$\beta_0$} &   \multicolumn{3}{c} {$\beta_1$}\\
\hline    $(\pd^a,~\pd^c)$                 &  $d_B$  & $d_{W1}$ & $d_{W2}$  & $d_B$    & $d_{W1}$ & $d_{W2}$  \\
\hline bottom left   & $0.16$  & $4.1$  & $0.53$ & $0.13$   & $5.8$ & $0.59$  \\
\hline bottom right & $0.18$  & $5.2$  & $0.67$ & $0.14$   & $5.6 $ & $0.50 $  \\
\hline top right       & $0.10$  & $5.1$  & $0.69$ & $0.12 $   & $5.5 $ & $0.51$  \\
\hline top left         & $0.11$  & $4.3$ & $0.53$  & $0.087$    & $5.1$ & $0.87 $  \\
\end{tabular}
\caption{Distances between the persistence diagram for the different parts of the force networks corresponding to 
Fig.~\ref{fig:ThreeNetworks}(a,c).}
\label{table:AC}
\end{table}

Consider again the networks shown Fig.~\ref{fig:ThreeNetworks}(a-c).  Divide each domain into 
four equal blocks: bottom left, bottom right, top right and top left.   Table~\ref{table:AB} gives the distances 
between corresponding blocks for the (a-b) $\pd$'s.
We see that the distances between the bottom left blocks are always larger than the distances between the 
other three blocks.  Hence we can conclude that  differences between the networks $\fn_I^a$ and $\fn_I^b$ are 
concentrated in the bottom left corner. Table~\ref{table:AC} shows the distances between the four blocks for the 
networks $\fn_I^a$ and $\fn_I^c$.  Here, the values  are very similar for all the blocks. This finding indicates that 
the changes are  distributed evenly over the entire domain.

So far, we have applied the distance concept to discuss the differences between small number of force networks.
Now we will proceed to analyze large number of force networks describing states of the systems as they are
evolve through the jamming transition.

\section{Results}
\label{sec:results}

In the DES that we have carried out, we slowly compress the system through a range of packing fractions $\rho = [0.63:0.90]$ 
as discussed in Sec.~\ref{sec:system}. While the system is being compressed, we output the force information at 
approximately fixed time intervals. In the present work we focus on results obtained by extracting approximately
$100$ samples, and averaged over $20$ realizations for the purpose of obtaining statistically significant results. 

We compute the distance between $\pd$'s computed using the force information from consecutive
samples. Since the $\pd$'s provide information about the geometry of the forces, it is useful
to consider the plots presented in this section as a measure of the rate of change of the geometry of the forces.   

In what follows, 
we focus first on the normal forces in a polydisperse frictional system ($r_p = 0.4,~\mu = 0.5$) and then proceed
to discuss the tangential forces in this system, as well as a polydisperse frictionless system ($r_p = 0.4,~\mu = 0)$.

\subsection{A polydisperse frictional system}

Figure~\ref{fig:mu05n} shows various  distances between  the $\pd$'s for consecutive (normal) force networks, averaged
over $20$ realizations for frictional polydisperse system ($r_p=0.4,~\mu = 0.5$).  We also plot the average  number of contacts 
per particle, $Z$.  The most pronounced feature is the 
dramatic change around the jamming point, $\rho_J$, loosely defined  as   the $\rho$ at which $Z\approx 3$~\cite{epl12}.  Each
distance, however, behaves differently as the system is compressed; as demonstrated below, these differences 
provide us with  additional insight regarding the evolution of forces  as the system goes through the jamming transition.

\begin{figure}
\centering
\subfigure[$\beta_0$.]{\includegraphics[width=3.1in]{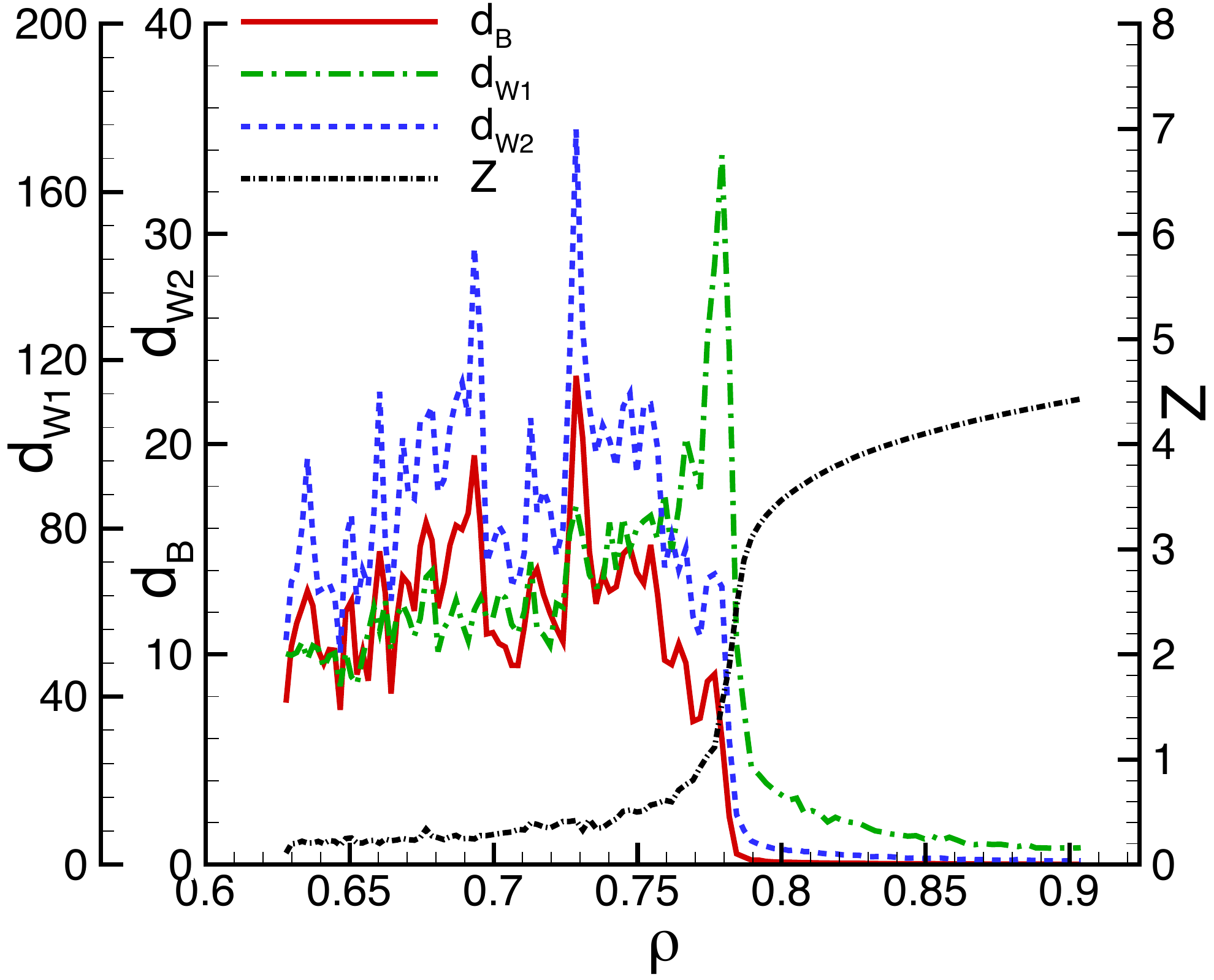}} \\
\subfigure[$\beta_1$. ]{\includegraphics[width = 3.1in]{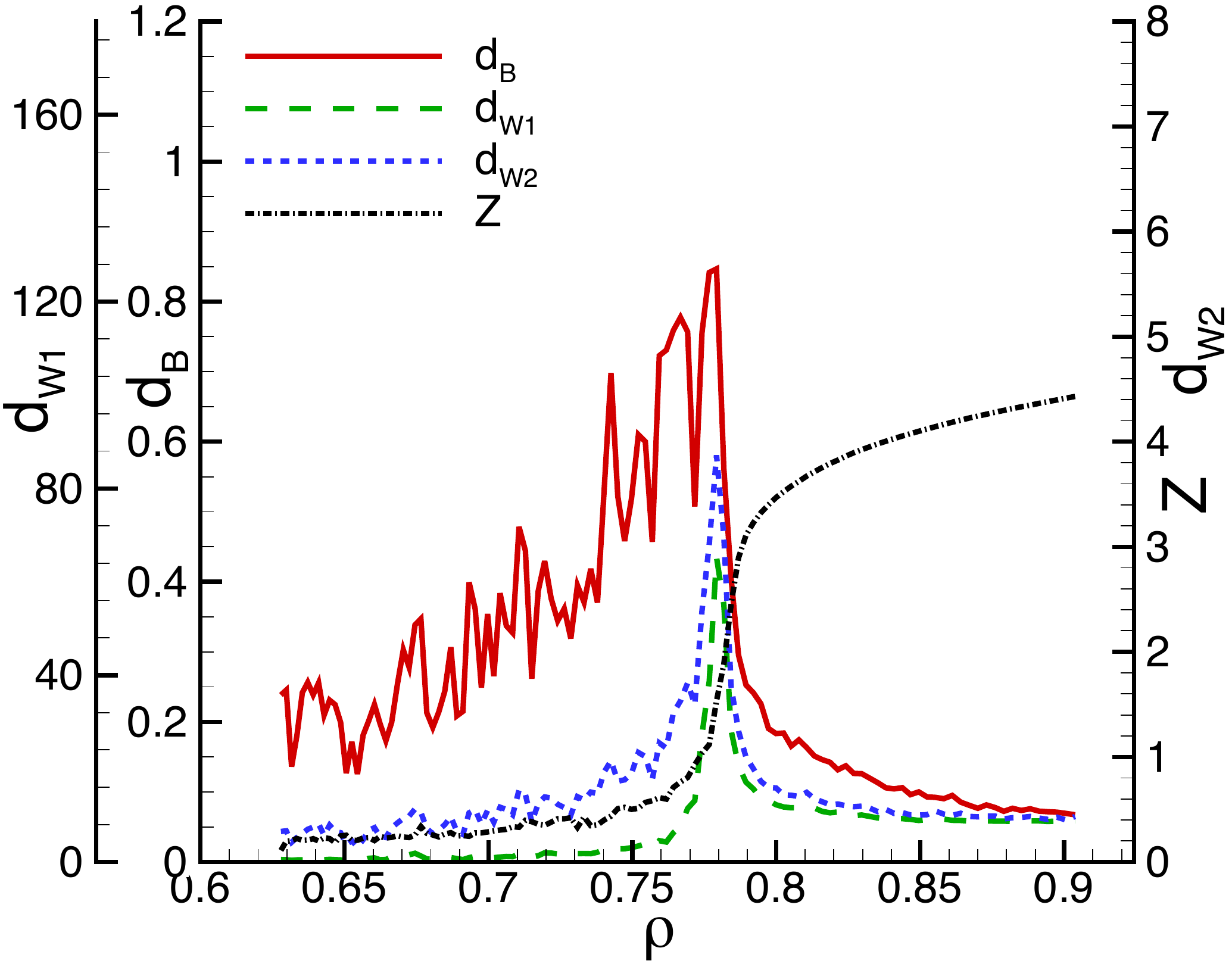}}
\caption{The distances between persistence diagrams for polydisperse frictional systems  ($r_p = 0.4,~\mu = 0.5$).  
The normal forces between particles are considered.
The $Z$ curve shows the average number of contacts per particle.   The results are obtained by averaging over $20$
realizations.
}
\label{fig:mu05n}
\end{figure}

Before comparing individual distances let us put their values into a context.   Recall that the  normal force used for computing $\pd$'s is 
normalized by the current average value. 
This average is  small  for $\rho < \rho_J$ and 
increases  (approximately linearly) after the jamming transition.  Using the landscape analogy there are at least two ways in which to 
interpret $d_B = 1$. The first is that
a geometric feature with lifespan equal to average force has either appeared or disappeared.  The second is that the
length of the lifespan of a single geometric feature has changed by the average force.

We begin our analysis by considering the evolution of the forces prior to jamming, $\rho < \rho_J$.
Figure~\ref{fig:mu05n}(a) gives the distances between consecutive $\pd_0$'s, 
therefore showing the rate of evolution of  connected components (force chains in the loose sense). The reported rate
clearly depends on the metric being used. The relatively large values ($d_B$ is an order of magnitude
greater than one for most of this regime) suggest rapid changes in the structure of the forces. There are
at least two types of events that are monitored; since there is still considerable room for particles
to move, we are observing the collisions of particles that create large spikes in the forces, and we are
observing changes in the connectivity of the system as particles come together, create a larger connected component  
with relatively strong force interactions inside this component, and then separate again. The $d_B$ metric is capturing 
the single largest of these  events 
while $d_{W^q}$, $q=1,2$, are measuring changes of the forces globally.  Figure~\ref{fig:mu05ratios}
plots the relative rates of change $d_B/d_{W^1}$ and $d_B/d_{W^2}$.  These relative rates of change are roughly
constant for $\rho < 0.75$.  Furthermore, $d_B/d_{W^2}\approx 0.75$ and  $d_B/d_{W^1}\approx 0.2$
suggesting that from frame to frame there is a single large event and a 
number of smaller  events  that contribute to the change in geometries.

\begin{figure}
\centering
\subfigure[$\beta_0$.]{\includegraphics[width=3.5in]{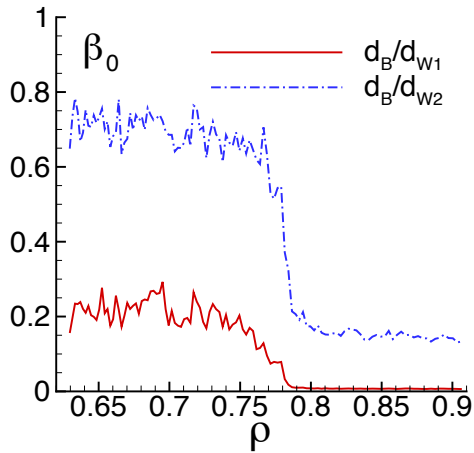}} \\
\subfigure[$\beta_1$. ]{\includegraphics[width = 3.5in]{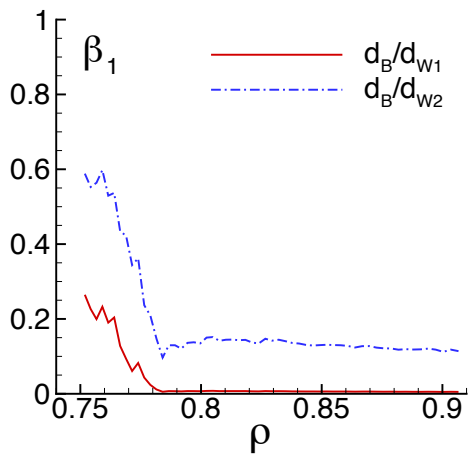}}
\caption{The ratios between the rates of change as measured by different metrics
for polydisperse frictional systems  ($r_p = 0.4,~\mu = 0.5$). 
}
\label{fig:mu05ratios}
\end{figure}

At $\rho \approx 0.75$, the behavior of the system begins to change. 
The rate of change as measured by $d_B$ decreases dramatically, suggesting absence of collisions
characterized by large relative velocity of the particles, or of 
large changes in connectivity
of the system accompanied by an increase of the forces acting between the colliding components.
For $0.75 < \rho < \rho_J$ the ratio of $d_B/d_{W^2}$ continues to be roughly constant until just before
the jamming  when it plunges.  This suggests that there continues to be a single large  event.
However, over this same range the rate of change as measured by $d_{W^1}$ dramatically increases
(Fig.~\ref{fig:mu05n}(a)) 
and hence the ratio of $d_B/d_{W^1}$ decreases, suggesting that the
number of small events is growing. The physical interpretation is that
as one approaches the jamming transition, there is a rapid reorganization of the structure of the forces,
taking place globally through many small rearrangements. 
 
Visual inspection of Fig.~\ref{fig:mu05n}(a-b) shows that there is a significant difference between the rate
of change in the geometries as measured by $\pd_0$ and $\pd_1$.  
For $\rho \ll \rho_J$, this is not surprising. The particles have room to move around. Thus, the
number of loops is small and furthermore, since it is unlikely that all the forces between the particles forming  loops are strong, 
the points in the $\pd_1$ tend to appear for smaller forces and have a shorter life span. 

With this in mind, it is perhaps not surprising that before jamming the rate of change of the loop structure generally increases as a function of $\rho$. At $\rho \approx 0.77$ the $d_B$ distance starts decreasing.
We interpret this to mean that if  the system is sufficiently packed then  it becomes difficult to support 
the appearance or disappearance of loops with large lifespans or for which the maximal or minimal forces change dramatically. 
It should be noted, however, that while the size of the individual changes becomes
constrained, the locations at which changes can occur do not.  
As it can be seen from Fig.~\ref{fig:mu05ratios}(b),
both ratios, $d_B/d_{W^2}$ and $d_B/d_{W^1}$, begin to decrease at $\rho \approx 0.76$.
The fact that $d_B/d_{W^2}$ decreases suggests that multiple large changes in the loop structures are occurring,
along with the increase in the number of small changes, as indicated by $d_B/d_{W^1}$.
We conclude therefore that during the jamming transition there is a
significant global reorganization of the loop structure with a variety of local changes of the magnitude comparable to the largest one. 

The analysis past the jamming transition, $\rho> \rho_J$, is simpler; the rate of change of the forces of 
the system slows dramatically. This is not surprising.  A large part of the contact network is fixed, thus
the particles cannot make and brake contacts, nor can the magnitudes of the normal forces change dramatically,
thus, for example, the rate of change measured by $d_B$ must be small.
What is more interesting to note is that the ratios of $d_B/d_{W^2}$ and $d_B/d_{W^1}$ remain essentially
constant throughout this regime and are roughly one and three orders of magnitude smaller, respectively,
than before the jamming transition.  This implies that even though the large events as measured by $d_B$
are getting smaller, they are also becoming more broadly distributed ($d_B/d_{W^2}$ smaller by one
order of magnitude) and
there are many more small events ($d_B/d_{W^2}$ smaller by three
orders of magnitude).

\subsection{Tangential forces}

While in the research related to force networks the focus is usually on normal forces, it is appropriate to ask whether evolution of 
tangential forces provides any additional information.   Figure~\ref{fig:mu05t} shows the corresponding results, where now the forces
are normalized with the average tangential force (as expected, the tangential force average is significantly smaller than the normal one).    
Perhaps the most interesting feature of the results shown in Fig.~\ref{fig:mu05t} is how similar they are to the results obtained
for normal forces, shown in Fig.~\ref{fig:mu05n} (except perhaps very close to $\rho_J$).   This finding suggests that both
for $\rho < \rho_J$ and for $\rho > \rho_J$, 
the evolution of tangential forces to large degree follows the evolution of normal forces, both regarding the connected components
and the loops.   For $\rho \approx \rho_J$, we see larger deviations between normal and tangential forces, with a particular
feature that the distances that include all the differences between the networks, such as $d_{W^1}$ and $d_{W^2}$, are significantly 
smaller for tangential forces, suggesting less dramatic evolution of these forces as the system goes through jamming transition.
The bottleneck distances, $d_B$, show however similar features and magnitudes for the normal and tangential forces throughout 
the evolution, including $\rho \approx \rho_J$, suggesting that the largest change in the tangential forces is slaved to the largest
change in the normal ones.

\begin{figure}
\centering
\subfigure[$\beta_0$.]{\includegraphics[width=3.1in]{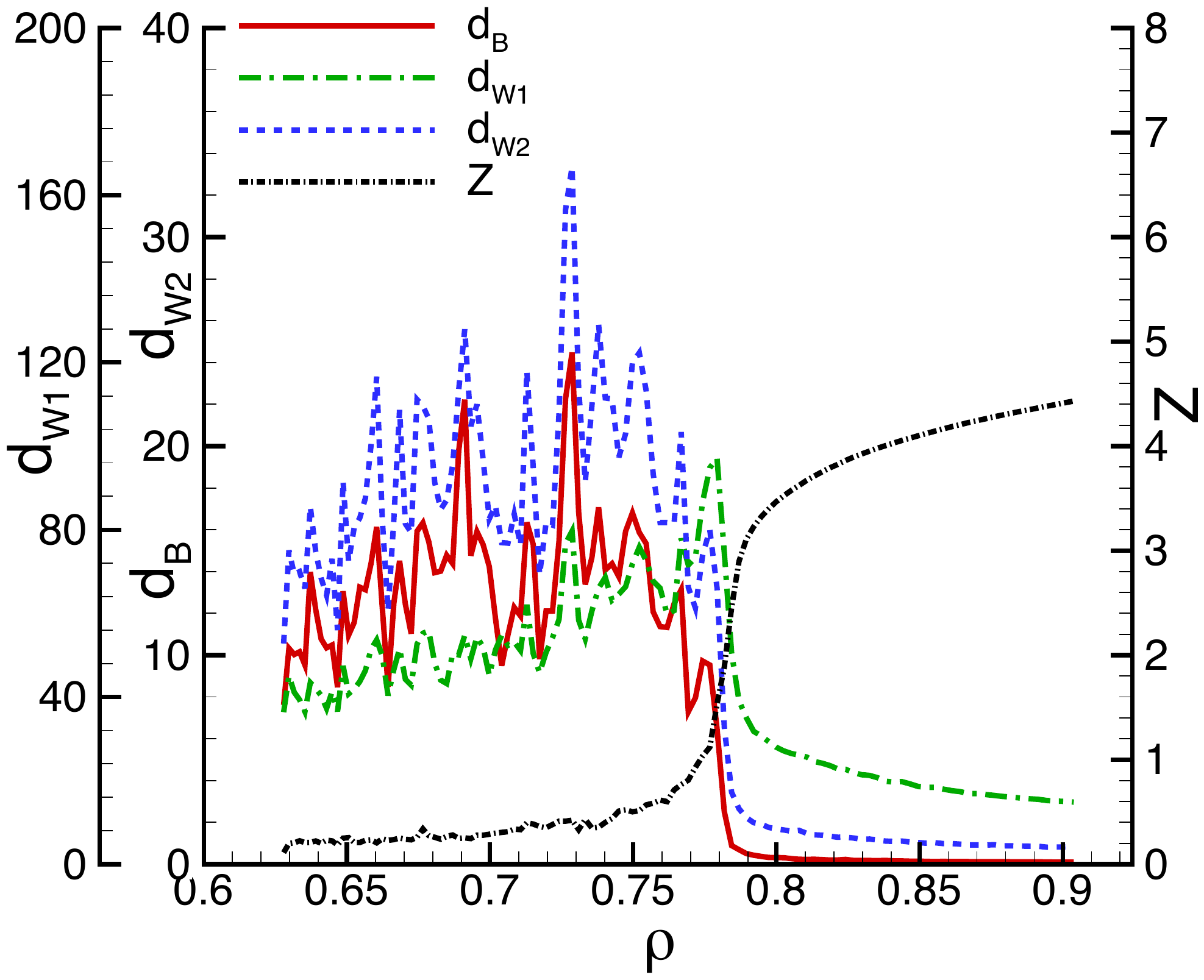}} \\
\subfigure[$\beta_1$. ]{\includegraphics[width = 3.1in]{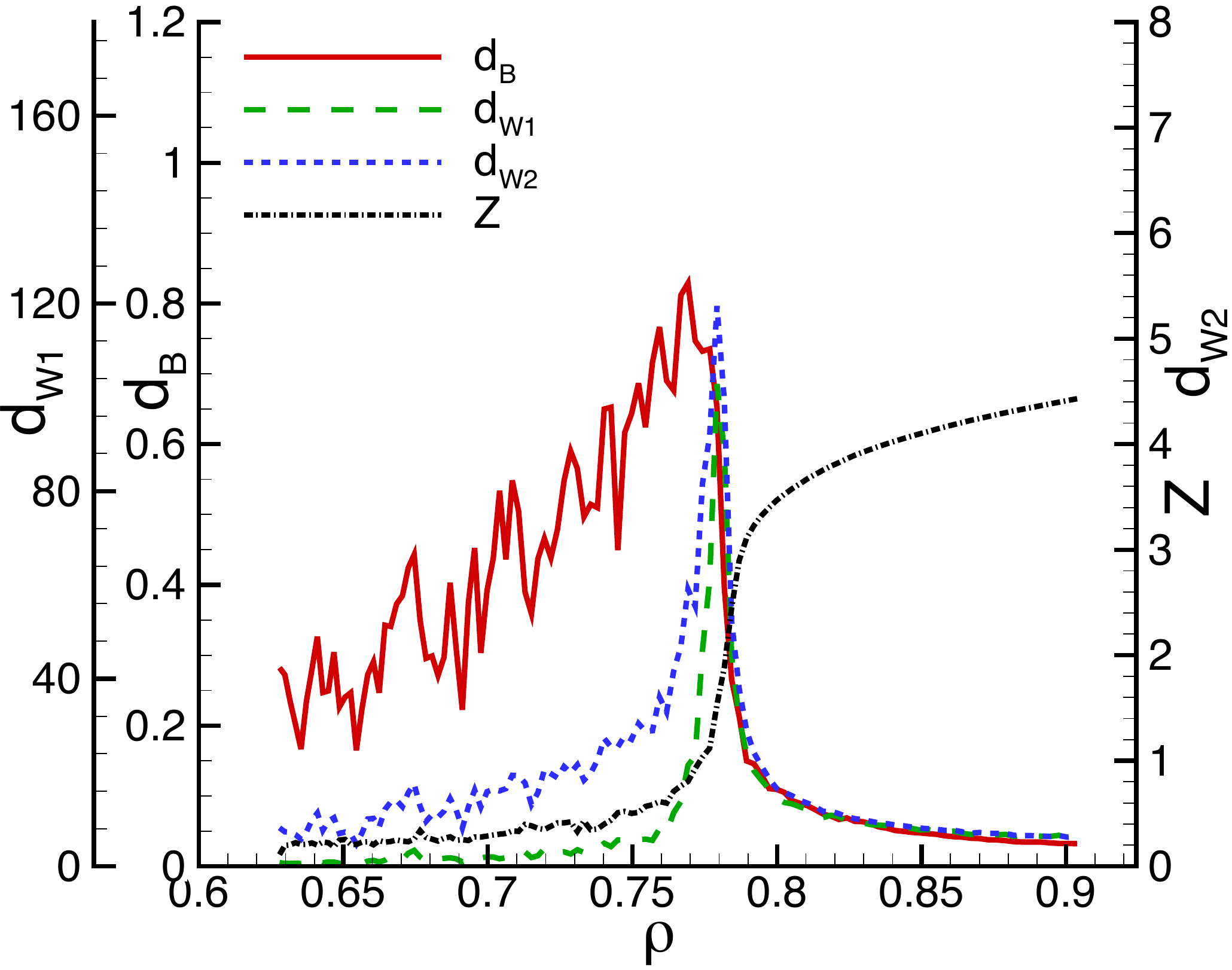}}
\caption{The distances between persistence diagrams for polydisperse frictional systems  ($r_p = 0.4,~\mu = 0.5$).  
The tangential forces between particles are considered.}
\label{fig:mu05t}
\end{figure}

\subsection{A polydisperse frictionless system}

Finally, we discuss briefly the normal force network for polydisperse frictionless ($\mu =0$)  system. 
In earlier work~\cite{pre13}, we discussed the information 
that can be obtained by considering the $\pd$'s, and found, based on the number of generators, 
that the force networks for $\mu = 0$ systems appeared to be extreme, in the sense
that the number of generators for frictionless system was significantly larger (see Fig.~6 in~\cite{pre13}), compared
to frictional systems.  
By considering
the distances between the states of the systems,  we can now discuss how friction influences the evolution of 
force networks. 

\begin{figure}
\centering
\subfigure[$\beta_0$.]{\includegraphics[width=3.1in]{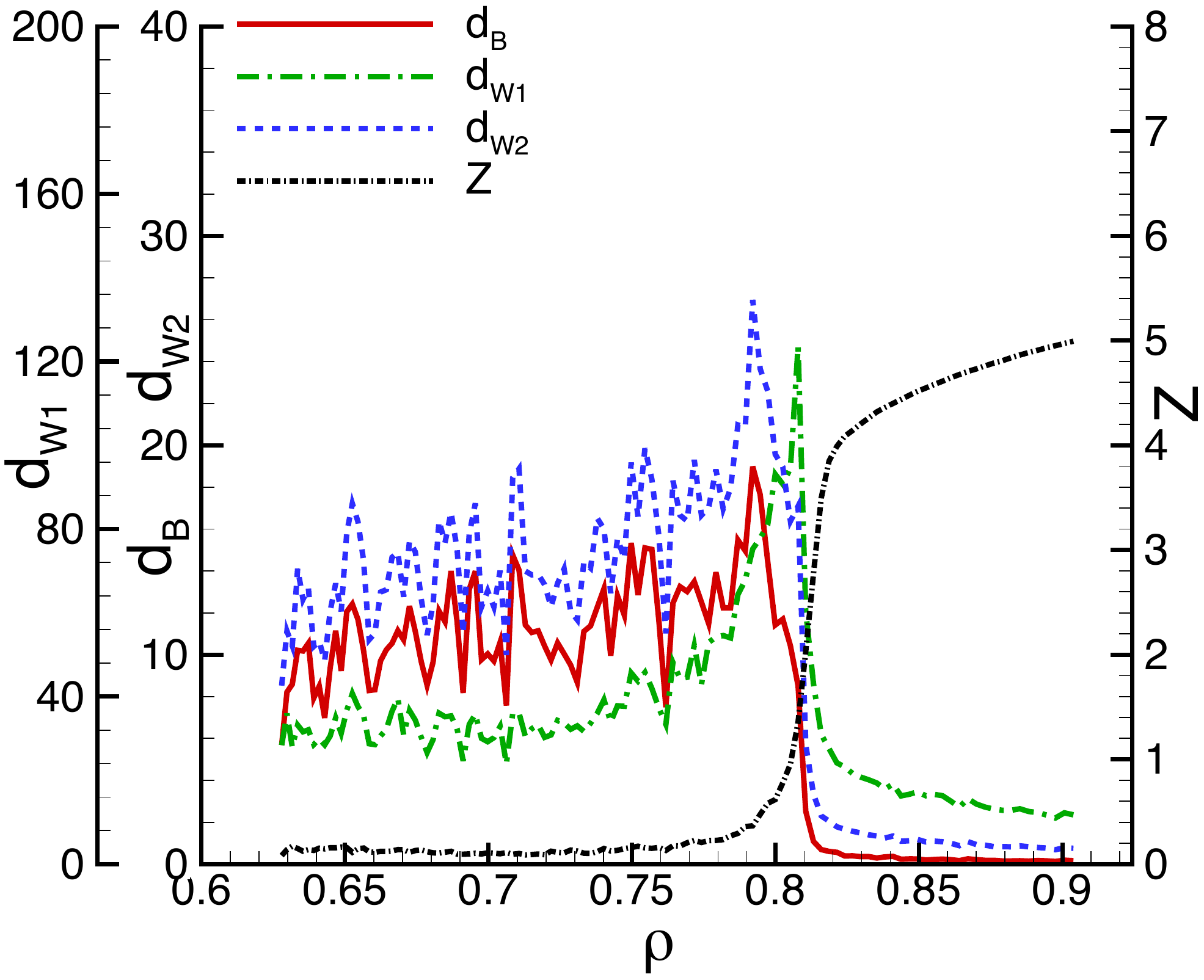}} \\
\subfigure[$\beta_1$. ]{\includegraphics[width = 3.1in]{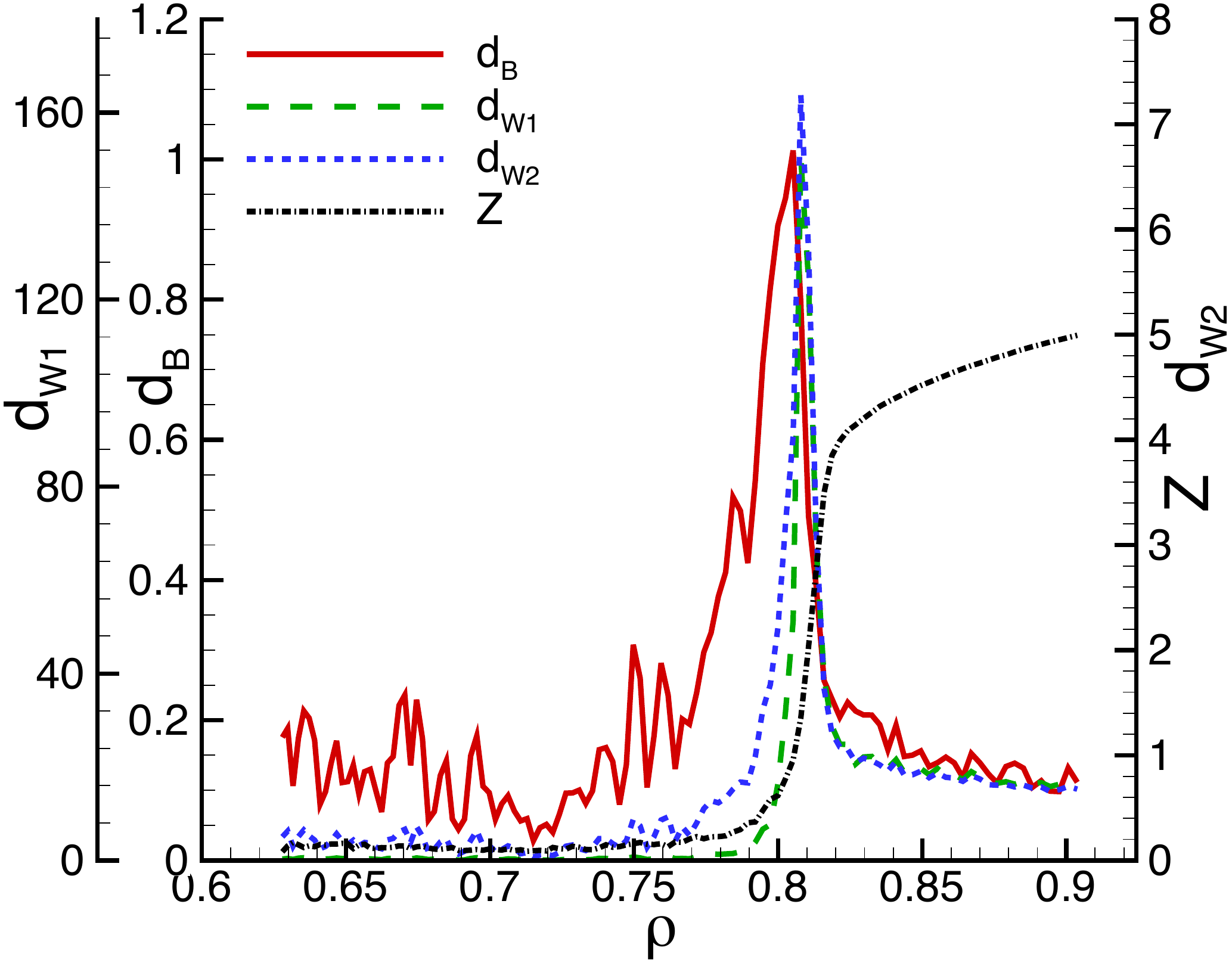}}
\caption{The distances between persistence diagrams for polydisperse frictionless systems  ($r_p = 0.4,~\mu = 0$).  
}
\label{fig:mu00}
\end{figure}

Figure~\ref{fig:mu00} shows the corresponding distances and the $Z$ curve.   Note that the jamming transition is shifted
to larger $\rho$'s for $\mu = 0$ systems~\cite{pre13}.   
Considering connected components, Fig.~\ref{fig:mu00}(a), we make two observations.
\\
$\bullet$
For  $\rho < 0.65$  the rate of change for $\mu =0$ systems, as measured by $d_b$ and $d_{W2}$, is roughly the same as for $\mu \neq 0$ systems. However, $d_{W1}$ is approximately $40$ \% smaller for $\mu = 0$.   In fact, as is seen
by comparing  Fig.~\ref{fig:mu00ratios}(a) and Fig.~\ref{fig:mu05ratios}(a)
over the entire range of $\rho$ adjusted for $\rho_J$, the ratio
$d_B/d_{W^1}$ is consistently larger for the $\mu = 0$ system while  $d_B/d_{W^2}$ is roughly the same. This suggests that
friction plays a role in the appearance of small events. 
\\
$\bullet$
Initially, as $\rho$ increases,
the rates of change increases more rapidly in the frictional system. However, both $d_B$ and $d_{W^2}$
begin slowing down at a much lower $\rho$ relative to $\rho_J$ in the frictional system, suggesting
that as the particles come in closer contact, friction limits the magnitude of the largest changes in the
geometry.

\begin{figure}
\centering
\subfigure[$\beta_0$.]{\includegraphics[width=3.5in]{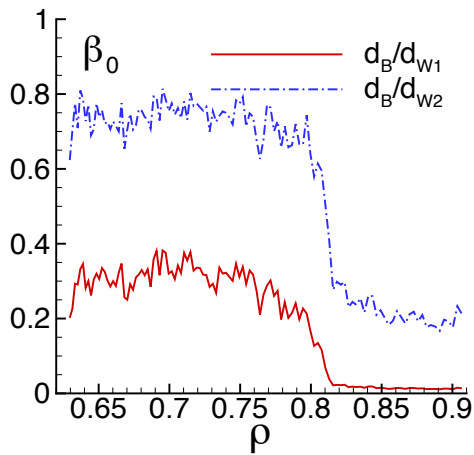}} \\
\subfigure[$\beta_1$. ]{\includegraphics[width = 3.5in]{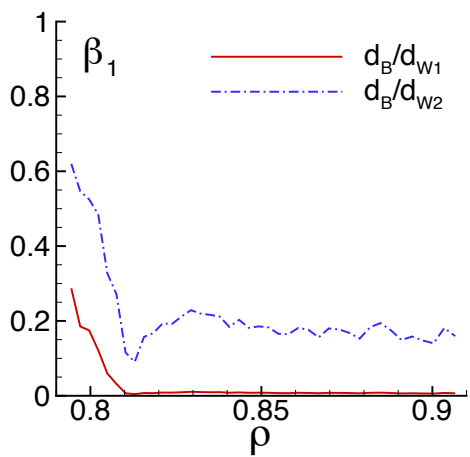}}
\caption{The ratios between the rates of change as measured by different metrics
for polydisperse frictional systems  ($r_p = 0.4,~\mu = 0$). 
}
\label{fig:mu00ratios}
\end{figure}

As indicated above, there are more points  in the $\pd_1$ for $\mu = 0$ than for
$\mu = 0.5$.  Thus the fact that for $\rho$ significantly less than $\rho_J$,
the rate of change is much less as measured by $d_B$ and $d_{W^2}$ in the $\mu = 0$ system as opposed to the
$\mu = 0.5$ system (compare Fig.~\ref{fig:mu00}(b) and Fig.~\ref{fig:mu05n}(b)), strongly suggests
that friction plays an important role in creation, destruction, and deformation of the loops when the system
is compressed but not yet jammed.  Observe that close to jamming, the rate at which the loop structure changes for 
the $\mu=0$ systems is significantly more dramatic, both regarding the largest change, measured by $d_B$, and regarding overall changes, measured by $d_{W1}$ and $d_{W2}$. Even after jamming, for $\rho> \rho_J$, 
the rate of change of the  $\mu = 0$ systems is approximately twice that of the $\mu = 0.5$ systems.   
It is interesting that the influence of friction for $\rho> \rho_J$ is much more obvious for loops than for connected components.
Further research is needed to better understand this finding.

\section{Conclusions} 
\label{sec:conclusions}

The evolution of force networks in dense particulate systems presents a complicated problem that is 
difficult to analyze using conventional techniques.   In this paper we show that computational 
homology can be used to address many 
aspects of the problem. In particular it provides a unique set of precise and well defined measures of the evolution of the geometry of the inter particle forces.
The main findings that apply to frictional and frictionless systems,
to normal and tangential forces, to both connected components (force
chains) and loops, and to all different distances between the considered states, are as follows:\\
$\bullet$  The comparison of considered measures describing the evolution of force networks shows that
for unjammed systems, large changes 
of the considered force networks are possible, but these changes typically
consist of local, isolated events.   However, as the system goes through the jamming transition, the measures
that we have implemented suggest large changes of the force networks on 
global, system-wide scale;\\
$\bullet$ The evolution of force networks is considerably different between unjammed, but dense systems, 
and the jammed ones.  For jammed systems, the evolution of the networks is significantly less dramatic, in the
sense that the distances between the consecutive states of the system are significantly smaller.  

In addition to these general findings, we also list the main findings that focus on the specific forces/systems/distances:\\
$\bullet$  The evolution of tangential forces is similar in its main features to the evolution of 
the normal ones, with some differences particularly close to $\rho_J$, where jamming occurs;\\
$\bullet$  There are significant differences in the evolution of force networks for frictionless versus frictional systems. 
With regard to the component structure of the force network (recall that this is related in a broad sense to 
so-called force chains) the peak rates of evolution before jamming are $20-40\%$ higher for the 
frictional system than for the frictionless system, but become quite similar after jamming. 
The differences are even more pronounced when the evolution of loop structures is considered. 
For $\rho \ll \rho_J$, the rate of evolution measured by $d_B$ and $d_{W^2}$ is larger for the frictional system. 
For $\rho \lessapprox \rho_J$, the relative rates change and thereafter the difference between the considered states are significantly more pronounced in frictionless systems.   

In this work, we have concentrated on describing the concept of distance between the persistence diagrams in 
the context of force networks, and have analyzed a particular set of systems (2D polydisperse circular particles
with/without frictional effects) exposed to a slow compression and sampled with prescribed sample rate. 
Furthermore, we have focussed on the generic, averaged features of the differences between considered force networks.
In future work, we will consider in more detail what is the influence of particle shape, sampling rate, and of the protocol used to evolve the considered particulate system, as well as analyze more
precisely the evolution of networks in single realizations~\cite{fast_sampling}.  

Before closing, we note that 
one particular strength of the computational homology approach is that it is system-independent
and can be applied to any particulate system, independent of the physical properties of the considered system, such as
the type of interaction between the particles, or dimensions of the physical space in which they live.   Therefore, the natural next step is to consider particulate systems characterized by different physical properties (cohesion, shape) or different  geometry (2D versus 3D).  Furthermore, the approach can be as well applied to experimental
systems, such as those built from photoelastic particles, where detailed information about the force networks is
available.    Analyses of such systems will be the subject of our future works.  

{\em Acknowledgments:}  
We thank Vidit Nanda for  valuable technical assistance with Perseus.  We acknowledge 
the support by NSF DMS-0835611, DTRA 1-10-1-0021 (LK); and NSF DMS-0915019, CBI-0835621
and contracts from DARPA  and AFOSR  (AG,MK,KM).

\appendix
\section {Discrete Element Simulations (DES)}
\label{app:DES}

In the simulations, circular grains
are confined to a square domain with rough walls composed of
monodisperse particles.  The walls move inward at constant speed,
$v_{c}$.  No
annealing of the system is carried out, and gravity is neglected.  
The particle-particle (and
particle-wall) interactions include normal and tangential components.
The normal force between particles $i$ and $j$ is
\begin{equation}
{\bf F}_{i,j}^n =
k_n x {\bf n} - \gamma_n \bar m {\bf v}_{i,j}^n~, 
\end{equation}
where $r_{i,j} = |{\bf r}_{i,j}|$, ${\bf r}_{i,j} = {\bf r}_i - {\bf r}_j$,
${\bf n} = {\bf r}_{i,j}/r_{i,j}$, and ${\bf v}_{i,j}^n$ is the
relative normal velocity.  The amount of compression is $x =
d_{i,j}-r_{i,j}$, where $d_{i,j} = {(d_i + d_j)/2}$, $d_{i}$ and
$d_{j}$ are the diameters of the particles $i$ and $j$. All quantities
are expressed using the average particle diameter, {\bf $d_{ave}$}, as
the lengthscale, the binary particle collision time $\tau_c = \pi
\sqrt{d_{ave}/(2 g k_n)}$ as the time scale, and the average particle mass,
$m$, as the mass scale.  $\bar m$ is the reduced mass, $k_n$ (in units
of ${ m g/d_{ave}}$) is the spring constant set to a value that
corresponds to that for photoelastic disks~\cite{geng_physicad03}, and
$\gamma_n$ is the damping coefficient~\cite{kondic_99}.  The
parameters entering the linear force model can be connected to
physical properties (Young modulus, Poisson ratio) as described {\it
 e.g.} in \cite{kondic_99}.

We implement the commonly used Cundall-Strack model for static
friction~\cite{cundall79}, where a tangential spring is introduced
between particles for each new contact that forms at time $t=t_0$.
Due to the relative motion of the particles, the spring length,
${\boldsymbol\xi}$ evolves as 
\begin{equation}
\boldsymbol\xi=\int_{t_0}^t {\bf v}_{i,j}^t~(t')~dt'~, 
\end{equation}
where ${\bf v}_{i,j}^{t}= {\bf v}_{i,j} - {\bf
 v}_{i,j}^n$.
For long lasting contacts, $\boldsymbol\xi$ may not
remain parallel to the current tangential direction defined by $\bf
{t}={\bf v}_{i,j}^t/|{\bf v}_{i,j}^t|$ (see,
e.g,.~\cite{brendel98}); we therefore define the corrected
\begin{equation}
\boldsymbol\xi{^\prime} = \boldsymbol\xi - \bf{n}(\bf{n} \cdot
\boldsymbol\xi)~,
\end{equation}
and introduce the test force 
\begin{equation}
{\bf F}^{t*} =
-k_t\boldsymbol\xi^\prime - \gamma_t \bar m {\bf v}_{i,j}^t~, 
\end{equation}
where $\gamma_t$ is the coefficient of viscous damping in the tangential
direction (with $\gamma_t = {\gamma_n}$).  To ensure that the
magnitude of the tangential force remains below the Coulomb threshold,
we constrain the tangential force as
\begin{equation}
{\bf F}^t = min(\mu_s |{\bf
 F}^n|,|{\bf F}^{t*}|){{\bf F}^{t*}/|{\bf F}^{t*}|}~, 
 \end{equation}
 and redefine
${\boldsymbol\xi}$ if appropriate.

For the initial configuration, particles are placed on a square
lattice and given random initial velocities; we have verified that the
results are independent of the distribution and magnitude of
these initial velocities.  The wall particles move at a uniform
(small) inward velocity $v_{c}= 2.5\cdot 10^{-5}$.  We integrate
Newton's equations of motion for both the translation and rotational
degrees of freedom using a $4$th order predictor-corrector method with
time step $\Delta t = {1/50}$.  In this work, we consider systems with  $N=2000$
particles with $k_n = 4\cdot 10^3$, $e_n =
0.5$, $\mu_s = 0.5$, and $k_t = 0.8
k_n$.

\bibliographystyle{apsrev}

\end{document}